\documentclass[letterpaper,twoside]{IEEEtran}
\usepackage{amsmath, amssymb, graphicx, psfrag, cite}
\graphicspath{ {./} }
\DeclareGraphicsExtensions{.pdf,.jpg}

\newtheorem{prop}{Proposition}
\newtheorem{thm}{Theorem}

\newtheorem{lemma}{Lemma}
\newtheorem{defin}{Definition}

\newcommand{\E}[1]{E[{#1}]}
\newcommand{\dc}{{d_{\tt{c}}}}
\newcommand{\dv}{{d_{\tt{v}}}}

\newcommand{\al}{\alpha}
\newcommand{\bt}{\beta}
\newcommand{\ep}{\varepsilon}
\newcommand{\dl}{\delta}

\DeclareMathOperator{\EE}{\mathbb{E}}

\title{Performance of LDPC Decoders with Missing Connections}%
\author{Linjia~Chang,~\IEEEmembership{Student Member,~IEEE,} 
	  Avhishek~Chatterjee,
	  and Lav R.~Varshney, \IEEEmembership{Senior~Member,~IEEE}%
\thanks{This work was presented in part at the 2016 IEEE International Symposium on Information Theory \cite{ChangCV2016}.}
\thanks{This work was supported in part by Systems on Nanoscale Information fabriCs (SONIC), one of the six SRC STARnet Centers, sponsored by MARCO and DARPA.}
\thanks{L.~Chang, A.~Chatterjee, and L.~R.~Varshney are with the Coordinated Science Laboratory, University of Illinois at Urbana-Champaign, Urbana, IL  61801 USA (e-mail: \{lchang10, avhishek, varshney\}@illinois.edu).}
}

\begin{document}
\maketitle

\begin{abstract}
Due to process variation in nanoscale manufacturing, there may be permanently missing connections in information processing hardware. Due to timing errors in circuits, there may be missed messages in intra-chip communications, equivalent to transiently missing connections. 
In this work, we investigate the performance of message-passing LDPC decoders in the presence of missing connections.  We prove concentration and convergence theorems that validate the use of density evolution performance analysis.
Arbitrarily small error probability is not possible with missing connections, but we find suitably defined decoding thresholds for communication systems with binary erasure channels under peeling decoding, as well as binary symmetric channels under Gallager A and B decoding. We see that decoding is robust to missing wires, as decoding thresholds degrade smoothly.
Moreover, there is a stochastic facilitation (SF) effect in Gallager B decoders with missing connections. We also conduct finite-length simulations, compare the decoding sensitivity to channel noise and to missing wiring, and perform preliminary error-tolerant manufacturing yield analysis.
\end{abstract}

\begin{IEEEkeywords}
Decoding,
error analysis,
message passing,
wiring,
stochastic facilitation
\end{IEEEkeywords}

\section{Introduction}
\label{sec:intro}

\IEEEPARstart{L}{ow-density} parity-check (LDPC) codes are prevalent due to their performance near the Shannon limit with message-passing decoders that have efficient implementation \cite{Gallager1963}. With the end of CMOS scaling nearing, there is interest in nanoscale circuit implementations of decoders, but this introduces concerns that process variation in manufacturing may lead to interconnect patterns different than designed \cite{AlYamaniRP2007,JengLW2007,CheeL2009}, especially under self-assembly \cite{HeathKSW1998,HaselmanH2010}. Yield on manufactured chips deemed perfectly operational is small---reports indicate $1$--$15$\% of circuit elements such as wires, switches, and transistors are defective \cite{HaselmanH2010}---leading to rather expensive industrial waste \cite{BreuerGM2004}. Changing the paradigm of circuit functionality from perfection to some small probability $\alpha$ of missing wires may eliminate much wastage and so it is of interest to characterize chips with permanently missing connections to determine suitable error tolerances.

Process variation in manufacturing also causes fluctuation in device geometries, which might prevent them from meeting timing constraints \cite{GhoshR2010},
especially in future nanoscale technologies like carbon nanotube circuits where device geometry control is especially difficult.
Such timing errors lead to missed messages in intra-chip communications, equivalent to transiently missing connections. Connections can also be missing transiently in programmable LDPC decoders \cite{MansourS2006}. It is thus also of interest to characterize decoders with transiently missing connections.

However, most fault-tolerant computing research assumes the circuit is constructed correctly and is concerned only with faults in computational elements. Peter Elias noted the following \cite{Elias1958}, but it remains true today: \begin{quote}J. Von Neumann has analyzed computers whose unreliable elements are majority organs---crude models of a neuron. Shannon and Moore have analyzed combinational circuits whose components are unreliable relays. Both papers assume that the wiring diagram is correctly drawn and correctly followed in construction, but that computation proper is performed only by unreliable elements. \end{quote} Such assumptions of fault-free circuit construction need to be reevaluated and performance analysis of computation with such wiring faults needs to be carried out.
 The only work we are aware of in fault-tolerant computing theory that briefly discusses wiring errors is \cite[Ch. 9.2]{WinogradC1963}.

We had previously extended the method of density evolution to decoders with faults in the computational elements and showed it is possible to communicate with arbitrarily small error probability with noisy Gaussian belief propagation \cite{Varshney2011}. Asymptotic characterizations were also determined for Gallager A \cite{Varshney2011} and Gallager B decoders with transient noise \cite{TabatabaeiYazdiCD2013,TabatabaeiYazdiHD2012,LeducPG2012}, energy optimization \cite{LeducPKG2015}, and both permanent and transient noise \cite{HuangLD2014}. Noisy decoding \cite{NgassaSDD2015, VasicC2007, DuprazDVS2015, RasheedIV2014, VasicIBR2015, Varshney2015b, BrkicIV2016}, and general noisy belief propagation, not necessarily in decoding \cite{HuangLD2015, KarbasiSSV2014}, have also been studied.
Recent studies show bit-flipping decoders with data-dependent gate failures can achieve zero error probability \cite{BrkicRIV2015, BrkicIV2016}, but with a subset of computation hardware that is reliable and no wiring diagram errors.

Rather than noise in computational elements, here we analyze the performance of message-passing decoders with missing connections and show that appropriately defined decoding thresholds are robust, in the sense of degrading smoothly.  This is true for both transiently and permanently missing connections in message-passing decoding circuits.  In certain settings, missing connections actually improve performance, resulting in stochastic facilitation (SF).\footnote{SF in decoding was observed with transient errors in computation, rather than with missing connections, initially in memory recall \cite{KarbasiSSV2014, ChenVV2014} and then in communications \cite{VasicIBR2015, IvanisVD2016}.}

A key difference between noisy computational elements and missing connections is that circuit technology enables 
detection of missing connections, see Sec.~\ref{sec:modelMiCo}.  This allows for simple adaptations of 
decoding algorithms, yielding better decoding performance 
under missing connections than under noisy components. A notable 
manifestation of this difference is in the so-called
decoder useful region. For transient or permanent noise, 
there is a strictly positive lower bound for the useful region, below which the channel output
is actually better than the decoded version since the internal decoder noise makes things worse.
For missing connections, there is no such lower boundary 
since the decoder asymptotically never degrades performance from the raw channel error rate.

The celebrated results of Richardson and Urbanke \cite{RichardsonU2001} developed density evolution for analyzing message-passing decoders for LDPC codes that are correctly wired.  Here we extend those results, so we can use the density evolution technique to characterize symbol error rate $P_e$, measuring the fraction of incorrectly decoded symbols at the end of message-passing decoding, even when the decoder has missing connections. We also show that the performance of decoders with transiently and with permanently missing connections are asymptotically equivalent. Traditionally \cite{RichardsonU2001}, there are thresholds for channel noise level $\ep$ below which $P_e$ can be driven to $0$ with increasing blocklength $n$. Unfortunately with missing connections in message-passing decoders, $P_e$ cannot be driven to $0$ in general without a significant modification of the decoding algorithm. Thus, following \cite{Varshney2011}, we let $\eta$ upper bound the final error probability achievable by decoders with missing connections after many iterations $\ell$ and give thresholds to $\ep$, below which ${\text{lim}}_{\ell \rightarrow \infty}P_e^{(\ell)} \le \eta$ under density evolution. 

We perform sensitivity analysis of density evolution to give insight into whether manufacturing or operational resources are 
more important in communication infrastructures.  We also comment on how our results inform semiconductor manufacturing yield analysis 
under the new paradigm of allowing some level of wiring error.  To demonstrate the practical utility of density evolution analysis, 
we also perform finite-length simulations of decoders with missing connections.

Sec.~\ref{sec:bg} discusses models of codes, channels, and LDPC decoders with both transiently and permanently missing connections, with a particular focus on hardware modeling. Sec.~\ref{sec:perf} develops tools including concentration and convergence theorems that provide validity to density evolution analysis. Secs~\ref{sec:peeling}, \ref{sec:ga}, and \ref{sec:gb} analyze the peeling decoder on the binary erasure channel (BEC) and the Gallager A and Gallager B decoders on binary symmetric channel (BSC) using density evolution, characterizing $P_e$ with missing connections. Sec.~\ref{sec:practice} connects our
work to practice through sensitivity analysis, finite-length simulations, and manufacturing yield analysis. Sec.~\ref{sec:clus} concludes by pointing out directions for further investigation.

\section{Background}
\label{sec:bg}
In this section we describe message-passing LDPC decoders with missing connections. We define the code and channels considered in this work and construct fault-free and missing-wire decoder models for characterization later. 
\begin{figure}
\centering
\includegraphics[width=2.8in]{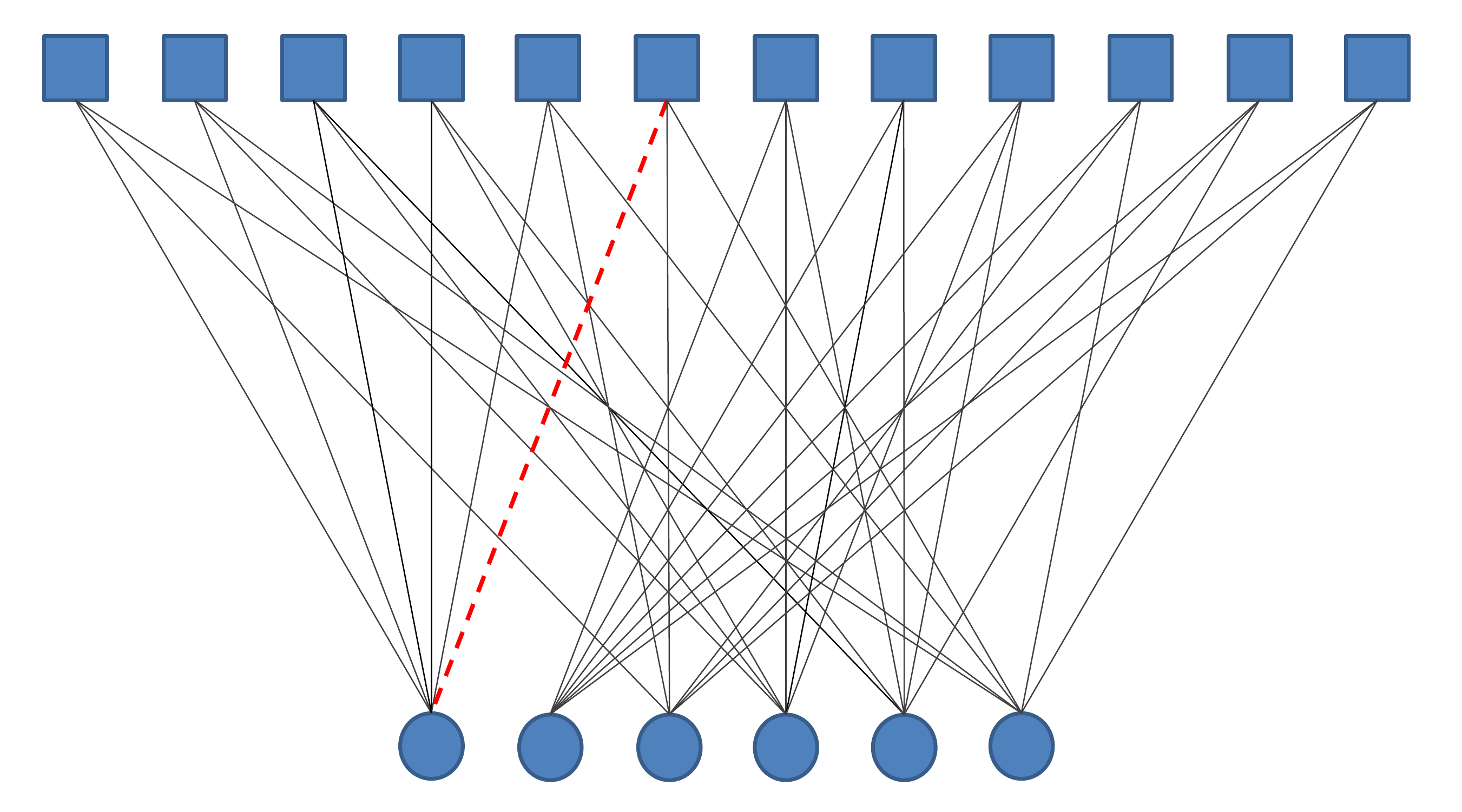}
\caption{Tanner graph of a $(3,6)$ regular LDPC code, with a missing wire for a corresponding message-passing decoder highlighted with a dashed line.}
\label{fig:tanner}
\end{figure}

\subsection{Ensemble of LDPC Codes and Channel}
We are concerned with the standard LDPC code ensemble $\emph{G}^n$, both regular and irregular.
First consider $(\dv,\dc)$-regular LDPC codes of length $n$, which can be defined by a bipartite Tanner graph with $n$ variable nodes of degree $\dv$ in one set, and $n\dv/\dc$ check nodes of degree $\dc$ in the other set (see Fig.~\ref{fig:tanner}).
For irregular codes $\emph{G}^n(\lambda,\rho)$, the degree distribution of variable and check nodes are denoted by functions $\lambda(x)=\sum_{d=2}^{\infty} \lambda_d x^{d-1}$ and $\rho(x)=\sum_{d=2}^{\infty} \rho_d x^{d-1}$, where $\lambda_d$ and $\rho_d$ specify the fraction of edges in the graph that are connected to nodes with degree $d$.
The variable nodes hold the codeword messages, and the check nodes enforce the constraints among variable nodes according to the code design. We consider this binary linear code ensemble as defined over the alphabet  
$\{\pm 1\}$. Although this section is general, for convenience, let us think of the communication channel as either BSC with output alphabet $\{\pm 1\}$ or BEC with output alphabet $\{\pm 1, ?\}$.

\subsection{Fault-Free Message-Passing Decoder}
The decoder operates by passing messages iteratively over the edges in the Tanner graph of the code. The implementation of such message-passing decoders in hardware follows the construction of the same Tanner graph too. We define a variable-to-check node message $u_{v\rightarrow c}$ and a check-to-variable node message $u_{c \rightarrow v}$. Message $u_{v'\rightarrow c'}$ from variable node $v'$ to check node $c'$ is often computed based on all incoming $u_{c\rightarrow v'}$ messages, where $c \in N(v')$ is a neighboring node of $v'$ and $c \neq c'$. For peeling, Gallager A, and Gallager B decoders, message $v_{c'\rightarrow v'}$ from check node $c'$ to variable node $v'$ is the product of all incoming $u_{v\rightarrow c'}$ messages, where $v \in N(c')$ is a neighboring node of $c'$ and $v \neq v'$.

\subsection{Missing Connections}
\label{sec:modelMiCo}
As discussed in Sec.~\ref{sec:intro}, there are two types of missing connections: permanent
missing connections caused by breaks in interconnects and transient missing connections
caused mainly by timing errors in intra-chip communication due to geometry variation in
circuitry.  Although specific statistical characterization is not reported in the semiconductor industry, process variation in manufacturing leads to both kinds of errors and can be fairly prevalent \cite{HaselmanH2010}.

For a given decoder circuit, permanent failure is modeled by removing each connection between variable and check nodes with probability $\alpha$ independently from others, before decoding starts.
These connections are never active once removed.
Conversations with circuit designers suggest that when an interconnect is broken in LDPC decoders implemented in a variety of device technologies, the measured signal voltage at this open-ended wire is neither low ($0$--$0.3V_{dd}$) nor high ($0.7V_{dd}$--$V_{dd}$); it is an intermediate floating value varying in the range ($0.3V_{dd}$--$0.7V_{dd}$) that can be differentiated from high/low values.  Hence we model it as an erasure symbol, ``?''.  

For the transiently missing connection setting, each connection between variable and check nodes is removed independently from others with probability $\alpha$ at each decoding iteration.
Transiently missing connections may occur due to timing error from incorrect geometry: 
consider a misalignment of synchronization when one branch of signal arrives after the computation at the
destination node has started, especially among those circuit implementations that do not
store the last signal sample. Transiently missing connections might similarly happen in programmable
LDPC decoder architectures \cite{MansourS2006}. Due to the difficulty in controlling device geometries, future carbon nanotube circuits are projected
to have a significant number of these transient missing connections.  Again we model as an erasure symbol, ``?''.

For notational convenience, let us restrict attention to decoders with messages in $\{\pm 1, ?\}$, but again concentration and convergence results demonstrated in Sec.~\ref{sec:perf} are general. Motivated by different concerns, \cite[Ex. 4.86]{RichardsonU2008} considered erasures in decoder messages as a representation of confidence, whereas \cite{MoriW2016} considered erasures as a way to capture check node or variable node failures in belief propagation.

\section{Performance Analysis Tools}
\label{sec:perf}
We now present mathematical tools to simplify the performance analysis of LDPC decoders with missing connections. In particular, we establish symmetry conditions for binary codes for easy analysis, and concentration and convergence results that endow the density evolution method with significance. Such results can be applied not only to decoders with binary messages, but also with larger message sets.

\subsection{Restriction to All-One Codeword}
Under certain symmetry conditions of the code, the communication channel, and the message-passing decoder, the probability of error is independent of the transmitted codeword.
\begin{itemize}
\item[C1.] {\bf Code Symmetry:} Code is a binary linear code.
\item[C2.] {\bf Channel Symmetry:} Channel is a binary memoryless symmetric channel \cite[Def.\ 4.3 and 4.8]{RichardsonU2008}.
\item[C3.] {\bf Check Node Symmetry:} If incoming messages of a check node are multiplied by $\{b_i \in \{\pm 1\}\}$, then the computed message is multiplied by $\prod_i b_i$.
\item[C4.] {\bf Variable Node Symmetry:} If the sign of each incoming message is flipped, the sign of the computed message is also flipped.
\end{itemize}
\begin{prop}
\label{prop:allone}
Under conditions C1--C4, in the presence of transiently or permanently missing connections, the probability of error of a message passing decoder is independent of the transmitted codeword.
\end{prop}
\begin{IEEEproof}
First consider mapping the erasure message ``?'', sent when a connection is missing, to $0$; thus the check-to-variable and variable-to-check messages are the messages computed at check node and variable node, respectively, multiplied by either $1$ (connection exists) or $0$ (missing connections).  It follows that messages passed between check and variable nodes satisfy the respective symmetry conditions \cite[Def.\ 4.82]{RichardsonU2008}. Hence, the result follows from \cite[Lem.\ 4.92]{RichardsonU2008}.
\end{IEEEproof}

In the sequel, we restrict the analysis of all models to the all-one codeword.

\subsection{Concentration around Ensemble Average}
\label{sec:conc}
We now show that the performance of LDPC codes decoded with missing-connection decoders stays close to the expected performance of the code ensemble for both transiently and permanently missing wires. The approach follows \cite{RichardsonU2001} and is based on constructing an exposure Martingale, obtaining bounded difference constants, and using Azuma's inequality.

Fix the number of decoding iterations at some finite $\ell$ and let $Z$ be the number of incorrect values held among all $\dv n$ edges at the end of the $\ell$th iteration for a specific choice of code, channel noise, and decoder with missing wires. Let $\E{Z}$ be the expectation of $Z$. Thm.~\ref{thm:conc} holds for decoders with both transiently and permanently missing connections.

\begin{thm}[Concentration around Expected Value]
\label{thm:conc}
There exists a positive constant $\bt = \bt(\dv,\dc,\ell)$ such that
for any $\ep > 0$, 
\[
\Pr[|Z-\E{Z}| > n \dv \ep / 2] \le 2e^{-\bt \ep^2 n} \mbox{.}
\]
\end{thm}
\begin{IEEEproof}
See Appendixes~\ref{app:concp} and \ref{app:conct} for permanent and transient missing connections, respectively.
\end{IEEEproof}
Recall Doob's Martingale construction from \cite{RichardsonU2001}, and the bounded difference constants for exposing channel noise realizations and the realized code connections, together with Azuma's inequality.  The main difference between the Martingale construction here and \cite{RichardsonU2001} is in the bounded differences due to the additional randomness from missing connections.

For permanently missing connections, one can think of the final connection graph being sampled from an ensemble of irregular random graphs with binomial degree distribution with average degrees $(1-\alpha)\dc$ and $(1-\alpha)\dv$, bounded by maximum degrees $\dc$ and $\dv$. Hence, the result follows from the result for correctly-wired irregular codes \cite{RichardsonU2001}. 

For transiently missing connections, the Martingale is constructed differently. Here instead of edges, for $\ell$ iterations, we sequentially expose the realization of edges at different iterations. Similar to \cite{Varshney2011} for transient noise in computational elements, the Martingale difference is bounded using the maximum number of edges over which a message can propagate in $\ell$ iterations, by unwrapping a computation tree. 

Note that $\beta$ is smaller for transient than permanent miswiring. The theorem extends directly to irregular LDPC codes.

\subsection{Cycle-Free Case}
We now show that the average performance of an LDPC code ensemble converges to an associated cycle-free tree structure, unwrapping a computation tree as in \cite{RichardsonU2001}.

For an edge whose connected neighborhood with depth $2\ell$ is cycle-free, let $q$ denote the expected number of incorrect values held along this edge at the end of $\ell$th decoding iteration. The expectation is taken over the choice of code, the messages received from the channel, and the realization of the decoder with missing wires. The theorems hold for both transiently and permanently missing connections.
\begin{thm}[Convergence to Cycle-Free Case]
\label{thm:tree}
There exists a positive constant $\gamma=\gamma(\dv, \dc, \ell)$ such that for any $\ep>0$ and $n>\frac{2\gamma}{\ep}$,
$$|\E{Z}-n\dv q|<n\dv\ep/2.$$
\end{thm}
\begin{IEEEproof}
The proof is identical to \cite[Thm. 2]{RichardsonU2001}, since introducing missing connections in a cycle-free tree structure does not change its cycle-free property.
\end{IEEEproof}
\begin{thm}[Concentration around Cycle-Free Case]\label{thm:treeconc}
There exist positive constants $\bt = \bt(\dv,\dc,\ell)$ and $\gamma=\gamma(\dv, \dc, \ell)$ such that for any $\ep>0$ and $n>\frac{2\gamma}{\ep}$,
$$\Pr[|Z-n\dv q|>n\dv \ep] \le 2e^{-\bt n\ep^2}.$$
\end{thm}
\begin{IEEEproof}
Follows directly from Thms.\,\ref{thm:conc} and \ref{thm:tree}.
\end{IEEEproof}
This concentration result holds for all message-passing decoders with missing connections.  In the sequel, we consider special cases of peeling, Gallager A, and Gallager B decoders.

\subsection{Density Evolution}
\label{sec:de}
With the concentration around the cycle-free case, it is clear that the symbol error rate $P_e$ of message-passing decoders with missing connections can be characterized with the density evolution technique. Let $P_e^{(\ell)}(g,\ep,\al)$ be the error probability of decoding a code $g \in \emph{G}^n$, after the $\ell$th iteration, where $\ep$ is the channel noise parameter, and $\al$ is the decoder missing wire probability. Density evolution evaluates the term:
$${\bar{P_e}}^{(\ell)}=\underset{n \rightarrow \infty}{\text{lim}}\E{P_e^{(\ell)}(g,\ep,\al)}.$$
The expectation is over the choice of code, channel noise realization, and missing wire realization.

Based on the proof of Thm.\,\ref{thm:tree}, we claim that the decoding error probability at any iteration $\ell$ for transiently and permanently missing connections, ${\bar{P_e}}^{(\ell)}_{T}$ and ${\bar{P_e}}^{(\ell)}_{P}$, become identical with the increase of the girth as blocklength $n$ increases. In particular, in density evolution the state variable $x_{\ell+1}$ is computed based on the $x_{\ell}$ values of nodes immediately below in the infinite tree. Each connection in the tree is encountered only once. In case of permanent failures each connection is present in the decoder with probability $1-\alpha$, whereas for transient failures
each connection is present at any iteration with probability $1-\alpha$. But, for
a given code symbol, its intrinsic messages traverse a particular edge only once if
the LDPC graph is a tree. Thus the messages experience the same statistical 
effect under permanent and transient failures. This results in the same probability
of error under both failures.

\begin{thm}
For any arbitrarily small $\delta = \delta(\dv,\dc,\ell) > 0$, $\sigma > 0$, and $\ell \ge 0$:
$$\Pr[|{\bar{P_e}}^{(\ell)}_{T}-{\bar{P_e}}^{(\ell)}_{P}| \ge \sigma] \le\delta.$$
\end{thm}
\begin{IEEEproof}
First, let $N_{\vec{e}}^{2\ell}$ be the neighborhood of an edge $\vec{e}$ with depth $2\ell$ in the decoding graph. Define the event $A_N$ as $N_{\vec{e}}^{2\ell}$ is not tree-like. It is shown that for a positive constant $\tau=\tau(\dv,\dc,\ell)$, $\Pr[A_N]\le \frac{\tau}{n}$ \cite[Thm.~2]{RichardsonU2001}. It implies the probability of exposing an edge multiple times decreases with increasing blocklength $n$ at any iteration $\ell$. Following the edge exposing procedure, ${\bar{P_e}}^{(\ell)}_{T}$ and ${\bar{P_e}}^{(\ell)}_{P}$ differ only when any edge $\vec{e}$ is exposed multiple times and the presence of $\vec{e}$ in the two decoding graphs with permanently and transiently missing connections differs.
Hence, $\Pr[|{\bar{P_e}}^{(\ell)}_{T}-{\bar{P_e}}^{(\ell)}_{P}| \ge \sigma]=\Pr[|{\bar{P_e}}^{(\ell)}_{T}-{\bar{P_e}}^{(\ell)}_{P}| \ge \sigma\big|A_N]\Pr[A_N]+\Pr[|{\bar{P_e}}^{(\ell)}_{T}-{\bar{P_e}}^{(\ell)}_{P}| \ge \sigma\big|{A_N}^c]\Pr[{A_N}^c]$.
Since $\Pr[|{\bar{P_e}}^{(\ell)}_{T}-{\bar{P_e}}^{(\ell)}_{P}| \ge \sigma\big|{A_N}^c]=0$, we can show $\Pr[|{\bar{P_e}}^{(\ell)}_{T}-{\bar{P_e}}^{(\ell)}_{P}| \ge \sigma] \le 1 \cdot \Pr[A_N]\le \frac{\tau}{n}$. As $n\to\infty$, this probability $\frac{\tau}{n}=\delta$ approaches $0$.
\end{IEEEproof}
In the sequel, no distinction 
is made between the analysis for transiently and permanently missing connection cases.

\subsection{Decoder Useful Region and Thresholds}
\label{sec:thresdef}
Usually density evolution converges to a certain stable fixed point with increasing number of iterations $\ell$. We define this fixed point as:
$$\bar{P_e}^{(\infty)}=\underset{\ell \rightarrow \infty}{\text{lim}}{\bar{P_e}}^{(\ell)}=\underset{\ell \rightarrow \infty}{\text{lim}}\underset{n \rightarrow \infty}{\text{lim}}\E{P_e^{(\ell)}(g,\ep,\al)}.$$

In order to decide when to use a decoder with missing connections, a \emph{useful decoder} is defined. A decoder is said to be useful and should be used instead of taking the codeword directly from the channel without decoding, if the asymptotic decoding error probability satisfies\cite{Varshney2011}: $$\bar{P_e}^{(\infty)}<\bar{P_e}^{(0)}=\ep.$$
The useful region of a decoder is defined as the set of parameters, in our case $(\ep, \al)$, that satisfies the above condition. 
Note that in case of transient computation
noise where computation is erroneous with probability $\tilde{\alpha}$ \cite{Varshney2011}, 
there are $(\ep, \tilde{\alpha})$ such that  $\bar{P_e}^{(\infty)}>\ep$. But under missing connections, for peeling, Gallager A, and Gallager B decoders, $\bar{P_e}^{(\infty)}\le\ep$ for any $(\ep, \al)$. This is 
because these decoders do not propagate erroneous messages under missing connections 
and hence cannot degrade symbols received from the channel.
When decoding with a fault-free decoder where $\al=0$, there exists an $\ep^*$ below which the final decoding error probability goes to 0 and a much larger value otherwise. We will see in the following sections that $\bar{P_e}^{(\infty)}$ does not go to zero for positive $\al$, but a threshold phenomenon still exists.\footnote{In general for $\bar{P_e}^{(\infty)}$ to go to 
$0$ in a faulty decoder, one needs to substantially change the decoder or 
to have a structural relationship between data and errors \cite{LeducPKG2015, BrkicRIV2015, BrkicIV2016}.} For every fixed $\al$, there exists a channel noise decoding threshold $\ep^*$, below which the final error probability $\bar{P_e}^{(\infty)}$ goes to a small value $\eta$. We call decoders that can achieve $\bar{P_e}^{(\infty)}$ that is lower than this small value $\eta$-reliable, and the channel noise level beyond which the decoder is $\eta$-reliable, the decoding threshold $\ep^*
$ \cite{Varshney2011}:
$$\ep^*(\eta, \al)=\text{sup}\big\{\ep \in [0,0.5] \big| \bar{P_e}^{(\infty)} \text{ exists and } \bar{P_e}^{(\infty)} <\eta\big\}.$$

\section{Peeling Decoder}
\label{sec:peeling}
Consider the peeling decoder for communication over a BEC with alphabet $\{\pm 1, ?\}$. The check node computation is a product of all messages $\pm 1$ it receives from neighboring variable nodes if none is ``?'', otherwise an erasure symbol ``?'' is sent. The variable node computation is to send any $\pm 1$ symbol received either from the other check nodes or from the channel, otherwise send ``?''. When the connection between two nodes is missing, the message exchanged is equivalent to ``?'', so peeling extends naturally to decoders with missing connections.  Note that this decoder satisfies the symmetry conditions C1--C4, so we can use density evolution assuming the all-one codeword was transmitted.

Although high-level intuition would suggest that the performance of decoding would degrade for any code and any decoder with missing connections, this is not the case 
as we later show for the Gallager B decoder. For the peeling decoder, the intuition holds and can be formalized using coupling techniques and
the fact that peeling decoders never propagate erroneous messages.
\begin{lemma}
\label{lem:BECmonoAlpha}
For any LDPC code $g$ 
with an arbitrary but finite blocklength, after a finite number of decoding iterations $\ell$, for both permanently and transiently missing connections, the symbol error 
probability $P_e^{(\ell)}(g,\ep,\al)$ increases monotonically with $\al$ for a 
given $\ep$.
\end{lemma}
\begin{IEEEproof}
The proof for monotonicity of $P_e^{(\ell)}(g,\ep,\al)$ follows by simple coupling arguments. For a specific LDPC code, consider two different missing connection probabilities $\al_1$ and $\al_2$, where $\al_1<\al_2$. Then, we couple the two missing connection processes as follows. Remove the wires with probability $\al_1$, and from this check-variable connection graph, remove each of the remaining connections with probability $\al_2-\al_1$. This gives a second missing connection process. It can be checked that the probability of missing connection in the second process is $\al_2$. Thus we can couple the missing connection processes to get a sample path dominance of connections. In this coupling, any realization of $\al_2$ process has more missing connections than that of $\al_1$. 

Now consider the probability of correctly decoding any bit $i$. Note that in peeling decoders, no erroneous messages are exchanged between check and variable nodes; only correct messages and erasures are passed along wires. A variable node $v_i$ holding message bit $i$ can be decoded correctly if either the received bit is correct, or the received bit is an erasure but $v_i$ receives a correct message through a path on the computation tree passing through one of its check nodes. The probability the received bit is correct is the same in case of both $\alpha_1$ and $\alpha_2$. So, let us compare the other probability. Now, by coupling as any realization of $\alpha_1$ has more connections than $\alpha_2$, if a correct message reaches $i$ following a path in the $\alpha_2$ graph, then that path also exists in the $\alpha_1$ graph. Thus, the event of receiving a correct message in case of $\alpha_2$ is a subset of that of $\alpha_1$. This proves monotonicity of correct decoding probability and so missing connections only degrade performance.
\end{IEEEproof}
A similar coupling argument yields an ordering relationship with respect to channel erasure probability $\ep$ for a given $\alpha$.

\subsection{Density Evolution Equation}
First, recall that the peeling decoding algorithm allows $\{\pm1,?\}$ to be sent, where ``?'' stands for an erasure caused by either the channel noise or a missing connection. In this case, the decoder only outputs either the correct message or an erasure symbol.

Consider a regular $(\dv, \dc)$ LDPC code, BEC channel with parameter $\ep$, and each wire independently disconnected with probability $\al$. Let $x_0,x_1,\ldots,x_{\ell}$ denote the fraction of erasures existing in the code at each decoding iteration.
The original received message from the channel is erased with probability $\ep$, so
${P_e}^{(0)}(\ep, \al)=x_0=\ep$.

Let $q_{in}$ be the probability that a node receives an erasure, and $q_{out}$ be the probability that a node sends out an erasure.
At a variable node, the probability that a given internal incident variable will be erased is the probability that both the external incident variable is erased and all other $\dv-1$ nodes are either disconnected or connected but erased.
\begin{align*}
q_{out} &= x_0\sum_{i=0}^{\dv-1} \binom{\dv-1}{i}{\al^i[q_{in}(1-\al)]^{(\dv-1)-i}}\\
&= \ep [\al+(1-\al)q_{in}]^{\dv-1}.
\end{align*}

At a check node, the probability that a given incident variable will not be erased is the probability that all $\dc-1$ other internal incident variables are not erased or disconnected. So the probability that a message is erased is
$$q_{out} = 1-[(1-q_{in})(1-\al)]^{\dc-1}.$$
Hence, the density evolution of the fraction of erasure between two consecutive decoding iterations is
$$x_{\ell+1}=\ep \big[\al+(1-\al)    \big( 1-[(1-x_{\ell})(1-\al)]^{\dc-1} \big)      \big]^{\dv-1}.$$

The density evolution result can be extended to irregular LDPC codes:
$$x_{\ell+1}=\ep \lambda\bigg(\al+(1-\al)\big(1-\rho[(1-x_{\ell})(1-\al)]\big)\bigg).$$
Let $f_{DE}(x_{\ell}, \ep, \al)=x_{\ell+1}$ be the recursive update function for the fraction of erasure, where $0\le\ep< 0.5$ and $0\le \al \le1$ is the domain of interest.

\subsection{Fixed Points}
The density evolution function $f_{DE}$ is non-decreasing in each of its arguments, given the other two. Thus, a monotonicity result similar to \cite[Lem.\ 3.54]{RichardsonU2008} holds. This also implies a convergence result for $x_{\ell}$, similar to \cite[Lem.\ 3.56]{RichardsonU2008}. So, for a given $\alpha$ and $\ep$, $x_{\ell}$ converges to the nearest fixed point of $x=f_{DE}(\ep,x,\alpha)$. Due to this existence of the fixed point, we can characterize the error probability when the decoding process is finished. The fixed points can be found by solving for the real solutions to the polynomial equation
\begin{equation}
\label{eq:fpz}
x-\ep \lambda\bigg(\al+(1-\al)\big(1-\rho[(1-x)(1-\al)]\big)\bigg)=0.
\end{equation}
We now prove that the decoding error probability is strictly positive by showing that $x=0$ is not a fixed point in \eqref{eq:fpz}.
\begin{lemma}
\label{lem:PbNot0}
For any irregular code ensemble $C^{\infty}(\lambda,\rho)$, there exists a $\delta>0$, such that the probability of error $P_e^{(\infty)}$ satisfies  $P_e^{(\infty)} > \ep\lambda(1-(1-\alpha)\rho(1-\alpha))>\delta>0$.
\end{lemma}
\begin{IEEEproof}
Since $x_{\ell}$ is monotonic, if $x_0 \le x_1$ then for any $\ell$, $x_{\ell+1} \ge x_{\ell} \ge x_{\ell-1}$. Now, for $x_0=0$, by substituting this value in $f_{DE}$,
$$x_1 =f_{DE}(0,\ep,\alpha)=\ep\lambda(1-(1-\alpha)\rho(1-\alpha))>0=x_0.$$
This implies that $\lim_{\ell \to \infty} x_{\ell}\ge f_{DE}(0,\ep,\alpha)$, for $x_0=0$.
But, as $x_{\ell}$ converges to the fixed point nearest to $x_0$ in the direction of monotonicity,
$x=0$ is not a fixed point and there is no fixed point in $(0,f_{DE}(0,\ep,\alpha))$ for any $\ep, \alpha>0$.
Thus we have $P_e^{(\infty)} >0$.
\end{IEEEproof}

Since this lemma shows all fixed points of the density evolution equation are
greater than $\ep\lambda(1-(1-\alpha)\rho(1-\alpha))$, decoding error probability cannot be
taken to zero. But this does not mean that the decoder is not useful. In fact it is always better 
to use the decoder, even when there are missing connections, rather than just taking corrupted symbols 
from the channel directly, since the peeling decoder never has incorrect messages. We can see this 
using the monotonicity of $f_{DE}(x,\ep,\al)$ in each of its arguments, given the other two.
For any channel and code, $x_0=\ep$, and it follows from $f_{DE}$ that $x_1 =f_{DE}(x_0,\ep.\al) \le \ep$. 
Hence $x_{\ell}\le \ep, \mbox{for all } \ell$, and $P_e^{(\infty)}\le \ep$. This is in sharp contrast to decoders with
computation noise, where decoder output can be strictly worse than channel output \cite{Varshney2011}.

\subsection{Performance Analysis}
In the previous section, we developed the recursive function to characterize the final error probability achieved by a peeling decoder with missing wires. Now we want to characterize the performance of such decoders.

For a peeling decoder, when $\ep=0$, the error probability stays at $0$ regardless of the quality of the decoder. When $\al=0$, it has been shown that there exists decoding threshold on the channel noise $\ep$, below which the final error probability can be driven to $0$ with the increase of decoding iterations \cite{RichardsonU2001}.
For the following analysis, we consider the system when $\ep >0$ and $\al>0$. Ideally, we want the error probability to be driven to $0$, but as demonstrated in Lem.\,\ref{lem:PbNot0}, this is impossible. Here we use the weaker notion of $\eta$-reliability defined in Sec.\,\ref{sec:thresdef}, where $\eta$ limits the final decoding error probability $P_e$.

\begin{figure}[!ht]
\centering
\includegraphics[width=3in, height=2.2in]{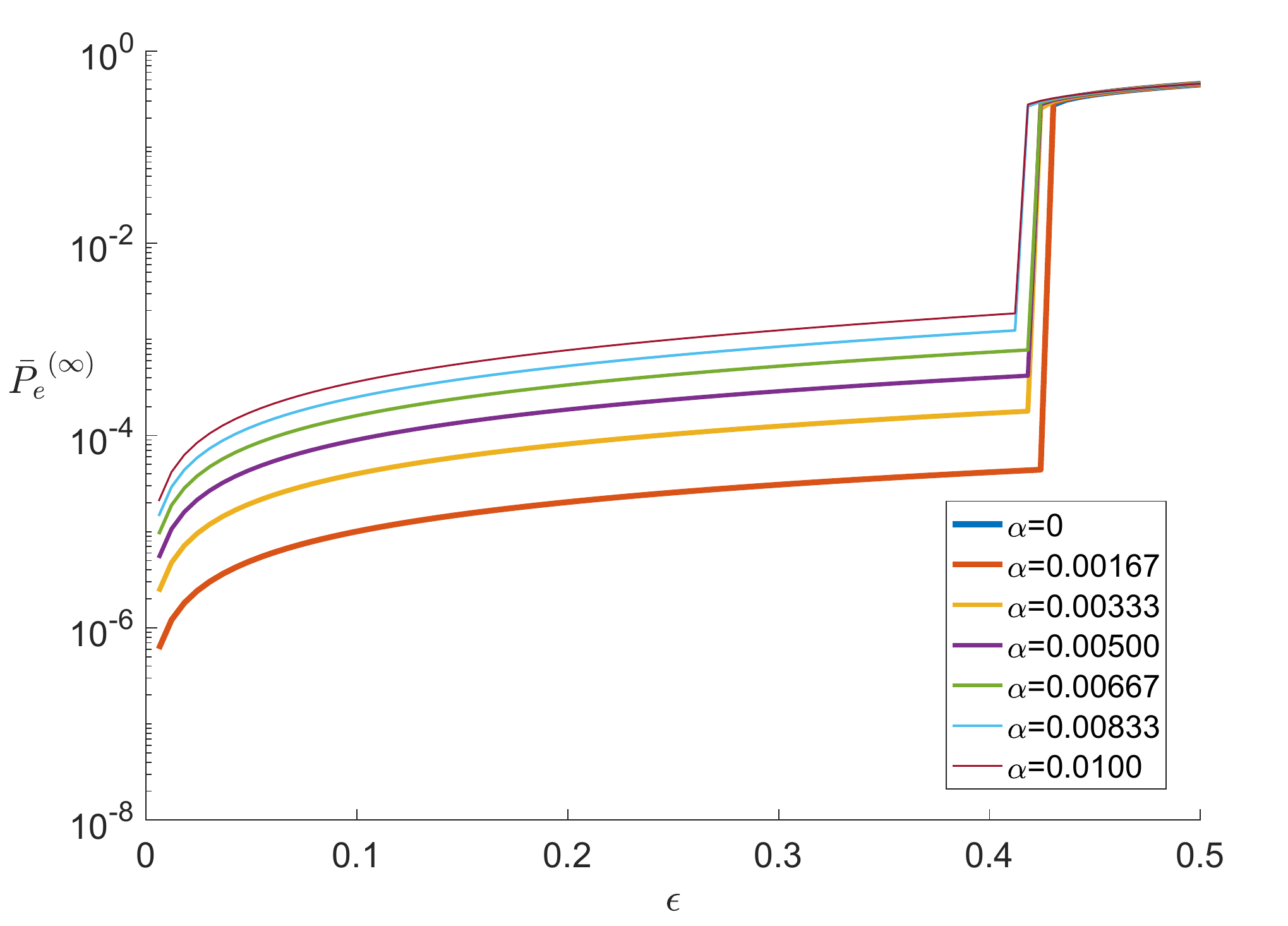}
\caption{Final symbol error rate of decoding a $C^\infty(3,6)$ LDPC code under peeling decoding algorithm with various missing connection probability $\alpha$ over BEC.}
\label{fig:etabec}
\end{figure}

\begin{figure}[!ht]
\centering
\includegraphics[width=3in, height=2.2in]{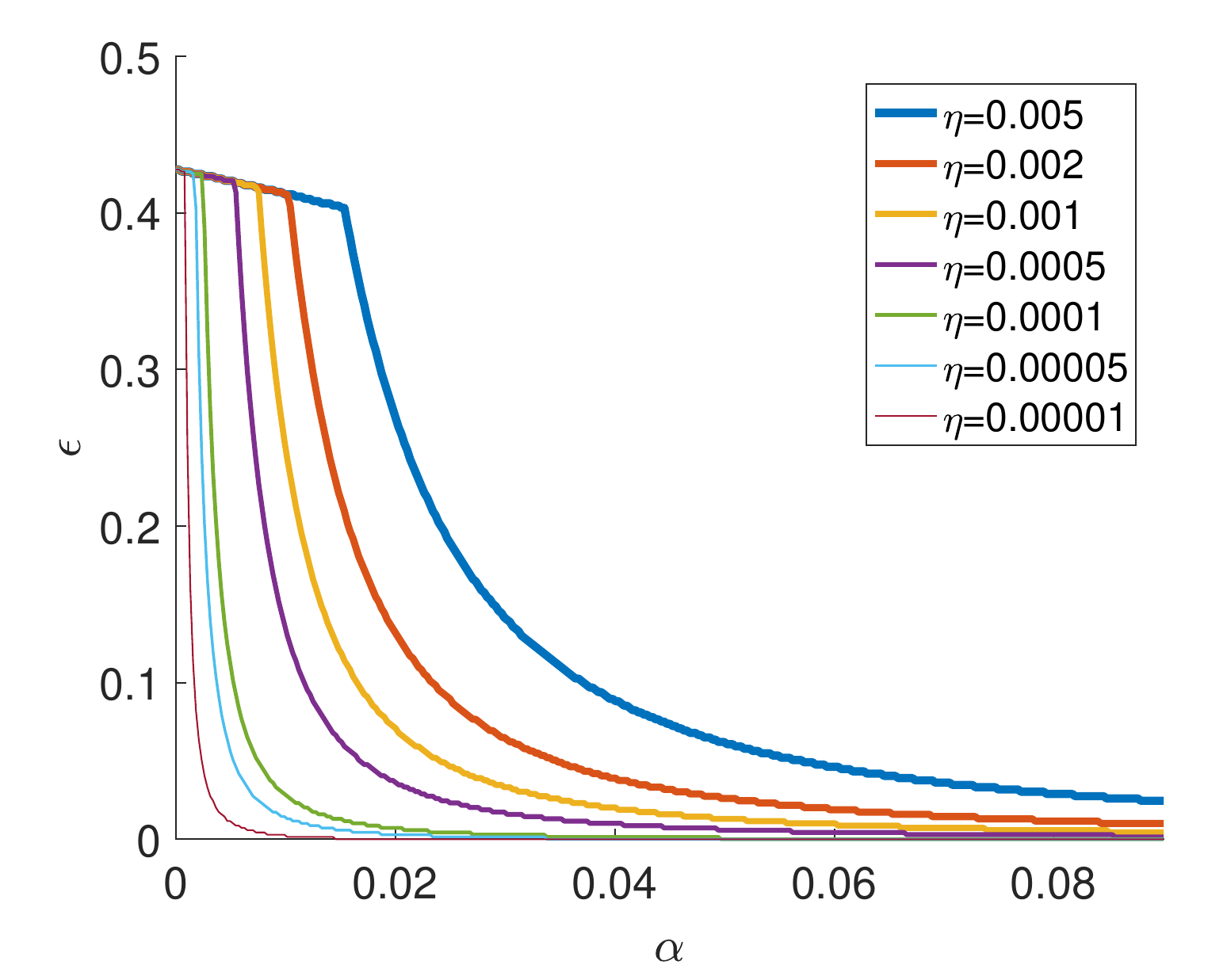}
\caption{Channel threshold of decoding a $C^\infty(3,6)$ LDPC code under peeling decoding algorithm over BEC for different given final error $\eta$-thresholds.}
\label{fig:thresbec}
\end{figure}

Fig.~\ref{fig:etabec} shows the final symbol error rate of decoding a $C^\infty(3,6)$ LDPC code under peeling decoding with various missing connection probabilities $\alpha$ over BEC($\ep$). It can be seen that given $\al$, there exists a threshold in channel noise level where a phase transition in $P_e$ happens. Fig.~\ref{fig:thresbec} illustrates such thresholds with the change of $\al$ under different small $\eta$-reliable constraints. 
An interesting phenomenon to notice in the decoding threshold is that there also exists a phase transition with the change of the decoder missing connection probability $\al$. With the increase of $\al$, for a fixed $\eta$-reliable decoder with missing connections, the decoding threshold first decreases linearly, and then exhibits more rapid decrease before convergence to zero.

\section{Gallager A Decoder}
\label{sec:ga}
Consider a fault-free Gallager A decoder for communication over a BSC. The messages are passed along the edges in the corresponding Tanner graph during decoding. A check node computes the product of incoming variable-to-check node messages $\{u_{v\rightarrow c}\}$; a variable node decides to flip the message from channel $y_v$ if all of the incoming check-to-variable node messages are $-y_v$ \cite{Gallager1963}.

With missing connections, the check node computation is not defined if an input is unknown (``?''). The product computed at the check node is the modulo-$2$ sum of all incoming messages to ensure that the parity constraints of the code are satisfied. When one of the bits involved in the parity is unknown, that parity check is no longer informative. This is because any bit of a linear code is equally likely to be $\pm 1$ (as complementing a binary codeword gives a codeword). So, for decoders with missing connections we make a natural adaptation: $u_{c\rightarrow v}= \mbox{``?''}$ if any of the incoming messages- is ``?''. We also make a natural adaptation for variable node computation: $-y_v$ is sent if more than one non-erasure check node messages are $-y_v$, and $y_v$ is sent otherwise. 

When it comes to Gallager A decoding over BSC, the messages being passed between nodes may carry erroneous information, unlike the peeling decoder, where the messages are either correct or erasure. So, for a sample path realization of channel and missing connections, a missing connection may prevent propagation of erroneous messages. Hence, unlike the peeling decoder, it is not apparent that there exists a stochastic dominance result like Lem.\,\ref{lem:BECmonoAlpha} between two different probabilities of missing connections. 
As fault-free decoding with the Gallager A algorithm satisfies conditions C1--C4, we can restrict analysis to the all-one codeword.

\subsection{Density Evolution Equation}
We find the probability for a variable node to compute $-1$  at iteration $\ell+1$, in terms of $x_{\ell}$. We consider a regular $(\dv, \dc)$ LDPC code and the adaptation of Gallager A decoding with erasure symbols for missing connections.

First note that since a BSC only outputs $\pm1$, a variable node never computes ``?'' with the Gallager A adaptation, even though it may receive (due to connection failure or check-node computes ``?'') or send the erasure symbol ``?'' (only due to connection failure).

The probability that a check node computation is $-1$ is:
\begin{align}
&\Pr\big\{\text{all } (\dc-1) \text{ variable nodes are connected and send odd} \nonumber \\
&\qquad \text{number of} -1\big\} \nonumber \\ 
& = (1-\alpha)^{\dc-1} \Pr\{\text{odd number of } (\dc-1)\text{ nodes send} -1\}
\nonumber \\
& = (1-\alpha)^{\dc-1} \tfrac{(1-(1-2x_{\ell})^{\dc-1})}{2}, \nonumber 
\end{align}  
where the last line follows using results from \cite[Sec.\ 4.3]{Gallager1963}.

The probability that a check node computation is $+1$ is:
\begin{align}
& \Pr\big\{\text{all } (\dc-1) \text{ variable nodes are connected and send even} \nonumber \\
& \qquad \text{number of} -1\big\} \nonumber \\ 
& = (1-\alpha)^{\dc-1} \Pr\{\text{even number of } (\dc-1) \text{ nodes have} -1\}
\nonumber \\
& = (1-\alpha)^{\dc-1} \tfrac{(1+(1-2x_{\ell})^{\dc-1})}{2}. \nonumber 
\end{align} 

The probability that a check-to-variable message is ``?'' is the complement of the probability that a check node computes $\pm 1$. Define $p_0$ to be
\begin{align} \label{eq:p0}
1-(1-\alpha)^{\dc-1}.
\end{align}

Consider a random variable $V\sim$ Binomial$(\dv-1,1-\alpha)$ with probability mass function $p_V(v)$, capturing the distribution of number of check nodes connected to a variable node.
Define $p_{+1}$ and $p_{-1}$ such that
\begin{align}&p_{+1}=(1-\alpha)^{\dc-1} \tfrac{(1+(1-2x_{\ell})^{\dc-1})}{2} \label{eq:p+}\end{align}
and
\begin{align}&p_{-1}=(1-\alpha)^{\dc-1} \tfrac{(1-(1-2x_{\ell})^{\dc-1})}{2} \label{eq:p-}.\end{align}

Now consider $x_{\ell+1}$, the error probability at a variable node at the $(\ell+1)$th iteration. The fraction of incorrect values held at this variable node is the sum of the probability of two events. The first event is that the message received from the channel is correct, and none of the incoming messages from the connected check nodes is correct, but not all of them are ``?'', and not only one says different while others are ``?''. The second event is that the message received from the channel is wrong, and at least one of the incoming messages from the connected check nodes is wrong or at most one check node is correct while all others are ``?''.

The probability of the first event is:
\begin{align*}
&\EE_V\bigg[(1-\ep)\big[\Pr\{\text{no connected check nodes sends } 1\}\\
&\text{ }-\Pr\{\text{all } V \text{ connected check nodes send } ``?"\}\\
&\text{ }-\Pr\{\text{one check node sends }-1\text{ while others send } ``?"\}     \big]     \bigg] \\
&=\sum_{v=1}^{\dv-1}p_V(v)(1-\ep)[(p_{-1}+p_0)^v-p_0^v-p_{-1}{p_0}^{v-1} ].
\end{align*}

The probability of the second event is:
\begin{align*}
&\EE_V \bigg[
\ep\big[
\Pr\{\text{at least one connected check nodes send }-1\}\\
& \text{ }+ \Pr\{\text{all } V \text{ connected check nodes send }``?"\}\\
& \text{ }+ \Pr\{\text{one check node sends }+1\text{ while others send } ``?"\}
\big]
\bigg]\\
&=\sum_{v=0}^{\dv-1}p_V(v)\ep    [1-  (p_{+1}+p_{0})^v +p_0^v +p_{+1}{p_0}^{v-1}].
\end{align*}

Let $x_{\ell+1}=f_{DE}(x_{\ell}, \ep, \al)$, and take the expectation of $V$ according to the binomial distribution to get
\begin{align*}
x_{\ell+1} =&f_{DE}(x_{\ell}, \ep, \al)\\
=& \ep\al^{\dv-1}+\sum_{v=1}^{\dv-1}\binom{\dv-1}{v}(1-\al)^v\al^{(\dv-1-v)}\\
&\cdot \bigg[   (1-\ep)[ (p_{-1}+p_0)^v-p_0^v-p_{-1}{p_0}^{v-1}] \\
& \text{  } +\ep [1-  (p_{+1}+p_{0})^v +p_0^v+p_{+1}{p_0}^{v-1}] \bigg].
\end{align*}

To extend to irregular LDPC ensembles, we take the average of the check node distribution and get:
\begin{align}p_{+1}^{(irr)}&=\rho(1-\al)\tfrac{1-\rho(1-2x_{\ell})}{2} \mbox{ and}  \label{eq:irr+} \\ 
p_{-1}^{(irr)}&=\rho(1-\al)\tfrac{1+\rho(1-2x_{\ell})}{2} \label{eq:irr-}.\end{align}
The terms in $f_{DE}(x_{\ell}, \ep, \al)$ have to be averaged over the variable node degree distribution of $\dv$ with function $\lambda(\cdot)$.

\subsection{Fixed Points}
It can be seen that $f_{DE}(x, \ep, \al)$ is monotonic in $x$ for a set of given $\alpha$ and $\ep$.  Hence, by the same arguments as for peeling decoders, for any initial $0\le \ep= x_0 \le 0.5$, $x_{\ell}$ converges to the nearest fixed point of the density evolution equation. We use $\tau_1 \le \tau_2 \le \tau_3 \le \cdots$ to denote these fixed points.

Note that for all $\ep >0, \al>0$, and $x_{\ell}=0$,  $f_{DE}(x_{\ell}, \ep, \al)=x_{\ell+1}>0$. This implies a result similar to Lem.\,\ref{lem:PbNot0} here. With the existence of channel noise and missing wiring, the decoding probability cannot be driven to 0. It is easy to show that for $\ep=0$, $f_{DE}(x, 0, \al)$ has one fixed point at $\tau_1=0$. We then focus on the case where $0<\ep<0.5$ for the following analysis.

Define $p^+(x)= (p_{-1}+p_0)^v-p_0^v-p_{-1}{p_0}^{v-1}$ and $p^-(x)=1-  (p_{+1}+p_{0})^v +p_0^v$.
An analytical expression for the channel threshold is the root ($\tau_2$) of the following expression between 0 and 0.5:
$$x\lambda(\al)+\lambda\bigg( p^+(x)-xp^+(x) + xp^-(x)\bigg)=x.$$

\begin{figure}
\centering
\includegraphics[width=3.5in, height=2.2in]{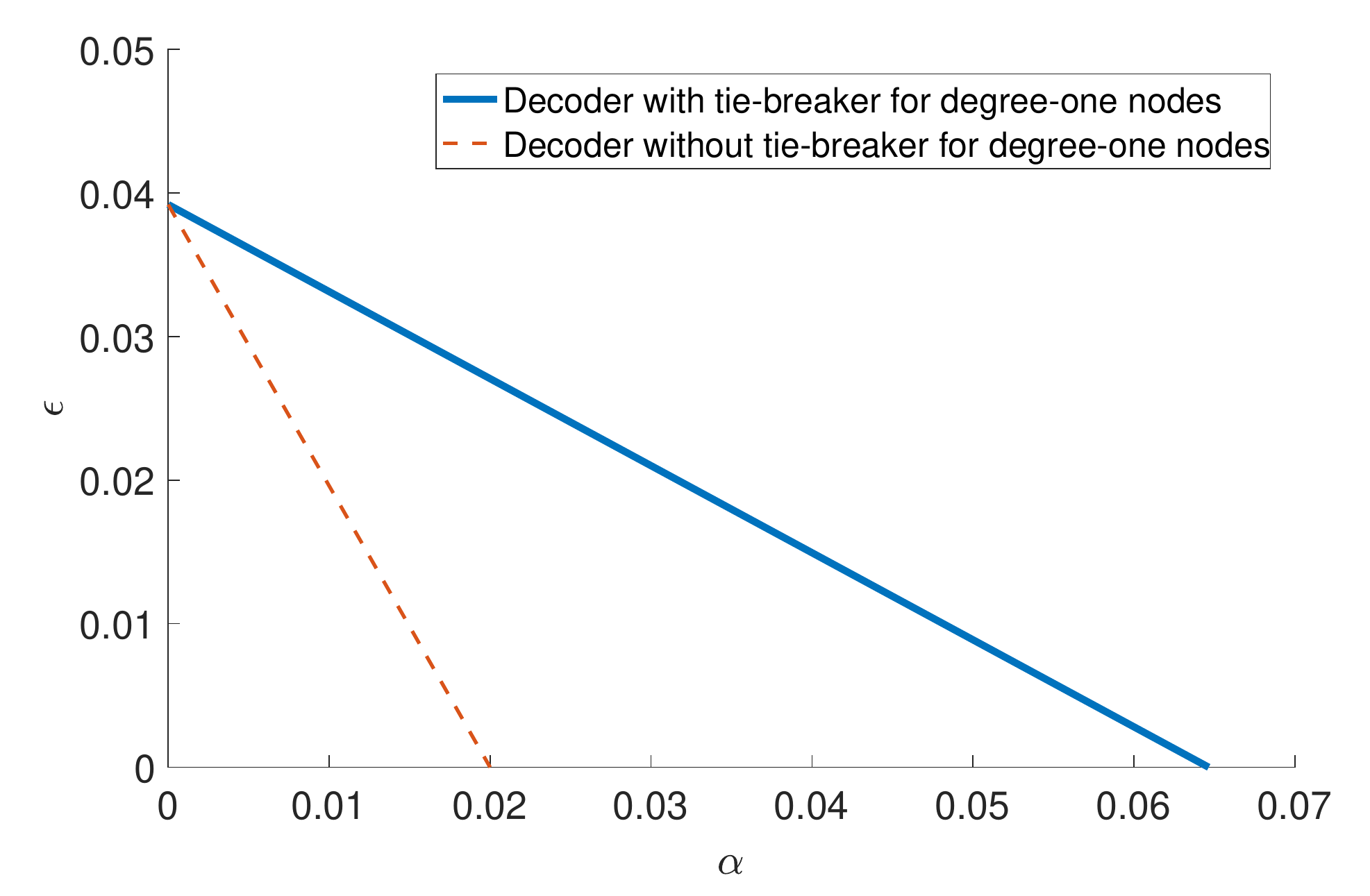}
\caption{Decoding a $C^\infty(3,6)$ regular LDPC code with $\alpha$-missing wire Gallager A decoding algorithm over BSC($\ep$). The useful region where it is beneficial to use decoder is between the curve and $\alpha$-axis.}
\label{fig:regbsc}
\end{figure}

The solid line in Fig.~\ref{fig:regbsc} shows the useful region of decoding for a $(3,6)$ regular LDPC code with missing connections, which is between $\tau_1$ and $\tau_2$ due to the monotonicity of function $f_{DE}$. Compared to \cite[Fig.~2]{Varshney2011} where computation at each node is noisy  with probability $\al$, the useful region of a decoder with missing connection is larger. In this case, decoders with missing connections outperform those with noisy computation. At any node, if the corresponding incoming message is missing rather than noisy with probability $\al$, the node is more likely to send a correct message than an erroneous one.

\subsection{Performance Analysis}
\begin{figure}
\centering
\includegraphics[width=3.5in, height=2in]{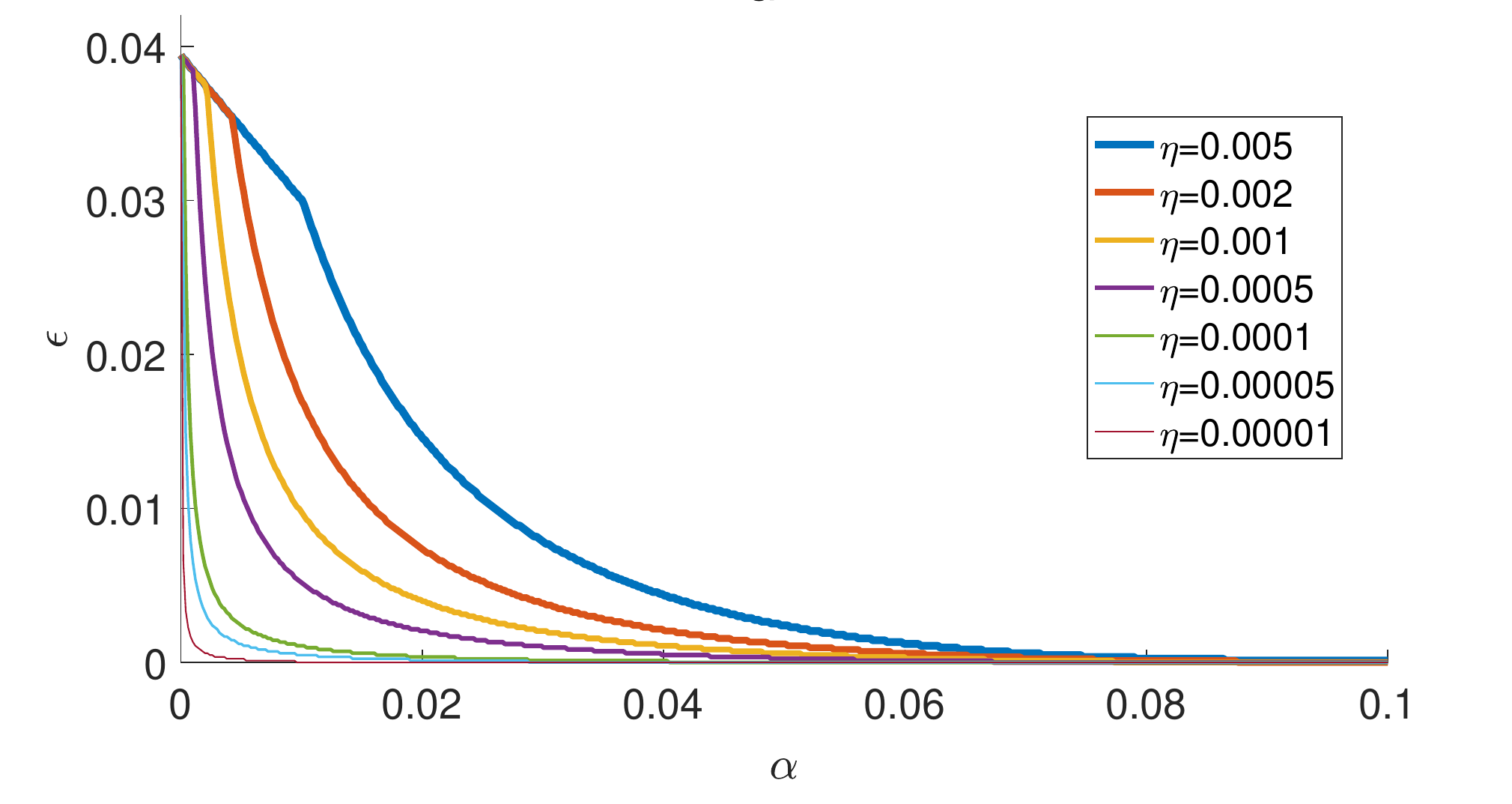}
\caption{$\eta$-thresholds for decoding a $C^\infty(3,6)$ regular LDPC code with $\alpha$-missing wire Gallager A decoding algorithm over BSC($\ep$).}
\label{fig:thresbsc}
\end{figure}

Fig.~\ref{fig:thresbsc} shows $\eta$-thresholds for communication over BSC($\ep$) with a Gallager A decoder with missing connections. Recall that for a $(3,6)$ regular LDPC code with a fault-free Gallager A decoder, the threshold is roughly $0.039$ \cite{BazziRU2004}. Note that $P_e$ can be driven to a fairly small number even with missing wires.  Decoding is robust to missing connection defects, though less than the peeling decoder over BEC.

As observed in Fig.\,\ref{fig:thresbsc}, a phase transition of the decoding threshold $\ep$ with the change of missing connection probability $\al$ noticed in the peeling decoder also exists here.
In contrast to classic settings, there may be degree-one nodes in decoding graphs due to the random missing connections. Hence, a tie-breaker at a variable node is necessary when the only incoming message from a check node is different from the received message from the channel. Since the channel message is more reliable than internal messages when there are missing connections in the decoder, we choose not to flip the channel message when the only incoming non-erasure message is the opposite. With this minor twist, the decoder useful region increases significantly, as shown in Fig.\,\ref{fig:regbsc}, where the dotted line shows the useful region of the decoder without the tie-breaker for degree-one case, choosing to flip the channel symbol when all incoming non-erasure messages are different from the channel symbol.

\section{Gallager B Decoder}
\label{sec:gb}
Gallager B decoders are usually more robust than Gallager A decoders without missing connections \cite{HuangLD2014}, so we modify the Gallager B algorithm by introducing erasure symbols for missing connections. In the Gallager B decoder, a check node performs the same operation with incoming variable-to-check node messages as Gallager A in Sec.\,\ref{sec:ga}, sending an unknown symbol ``?'' if one of the incoming messages is from a disconnected node. At a variable node however, instead of flipping the current value $u$ only when all the incoming messages from connected nodes say $-u$, a variable node in the Gallager B decoder decides to correct the current value $u$ when there are more than $b$ number of incoming messages that are $-u$. This threshold can be iteration-specific to reach optimality. Here, we fix the majority criterion, $b^*=\lfloor(\dv+1)/2 \rfloor$, in all iterations because this threshold results in small error probability independent of iteration number in fault-free Gallager B work \cite[Sec. 5]{Guruswami2006_arXiv}. We also choose $b^*$ based on the designed code without counting the number of actually connected nodes for simplicity, and it is verified numerically that there is no significant difference in performance.

Similar to the Gallager A model developed in Sec.\,\ref{sec:ga}, the codeword symmetry conditions C1--C4 are all satisfied in invoking Prop.\,\ref{prop:allone}.

\subsection{Density Evolution Equation}
The density evolution equation for the Gallager B decoder is similar to Gallager A. Consider a regular $(\dv, \dc)$ LDPC code and all-one codeword transmitted over BSC. At iteration $\ell$, the probability of a check-to-variable message is ``?'', $+1$ or $-1$ with probabilities $p_0$, $p_{+1}$ and $p_{-1}$, respectively, which have the same expressions as in
Sec.~\ref{sec:ga}.

Now consider $x_{\ell+1}$, the error probability at a variable node at the $(\ell+1)$th iteration. The fraction of incorrect values held at this variable node is the sum of the probability of two events. The first event is that the message received from the channel is correct, and at least $b=\lfloor(\dv+1)/2 \rfloor$ check nodes are connected and send incorrect messages. The second event is that the message received from the channel is wrong, and at most $b-1=\lfloor(\dv-1)/2 \rfloor$ of the incoming messages from the check nodes are correct. Consider a random variable $V\sim$ Binomial$(\dv-1,1-\alpha)$ capturing the distribution of the number of check nodes connected to a variable node.

The probability of the first event is:
\begin{align*}
&\EE_V\big[(1-\ep)\Pr\{\text{at least }b\text{ check nodes are connected}\\
&\qquad \text{and send}-1\} \big] \\
&=\sum_{v=b}^{\dv-1}p_V(v)(1-\ep){p_{-1}}^v{(1-p_{-1})}^{\dv-1-v}.
\end{align*}

The probability of the second event is:
\begin{align*}
&\EE_V\big[\ep\Pr\{\text{at most } (b-1) \text{ check nodes send}+1\}\big]\\
&=\EE_V\big[\ep[1-\Pr\{\text{at least }b\text{ check nodes are connected}\\
&\qquad \text{and send}+1\}]\big]\\
&=\sum_{v=b}^{\dv-1}p_V(v)\ep [1-  {p_{+1}}^v{(1-p_{+1})}^{\dv-1-v}].
\end{align*}
Taking the expectation of $V$ according to the binomial distribution, we have
\begin{align*}
x_{\ell+1} =& \sum_{v=b}^{\dv-1}\binom{\dv-1}{v}(1-\al)^v\al^{(\dv-1-v)}
\bigg[   (1-\ep)[ {p_{-1}}^v\\
&\cdot {(1-p_{-1})}^{\dv-1-v} ] + \ep    [1-  {p_{+1}}^v{(1-p_{+1})}^{\dv-1-v}] \bigg].
\end{align*}
The density evolution equation can also be extended to irregular LDPC codes, with changes in parameters $b(x)=\lfloor(\lambda(x)+1)/2 \rfloor$, $p_{+1}^{(irr)}$, and $ p_{-1}^{(irr)}$ defined in expressions (\ref{eq:irr+}) and (\ref{eq:irr-}).

\subsection{Performance Analysis}
We carry out detailed  performance characterization of a Gallager B decoder with missing connections and show that such a decoder is indeed more robust to missing connections than Gallager A.

Note when variable node degree $\dv=3$ for a regular LDPC code, a fault-free Gallager B decoder with the defined threshold $b=\lfloor(\dv+1)/2 \rfloor$ is equivalent to a fault-free Gallager A decoder. However, due to the modification the of Gallager A decoder to keep the received channel message when there exists only one incoming message, in the case of missing connections, these two decoders behave differently for decoding a $C^\infty(3,6)$ regular LDPC code. 
\begin{figure}
\centering
\includegraphics[width=3.5in, height=2in]{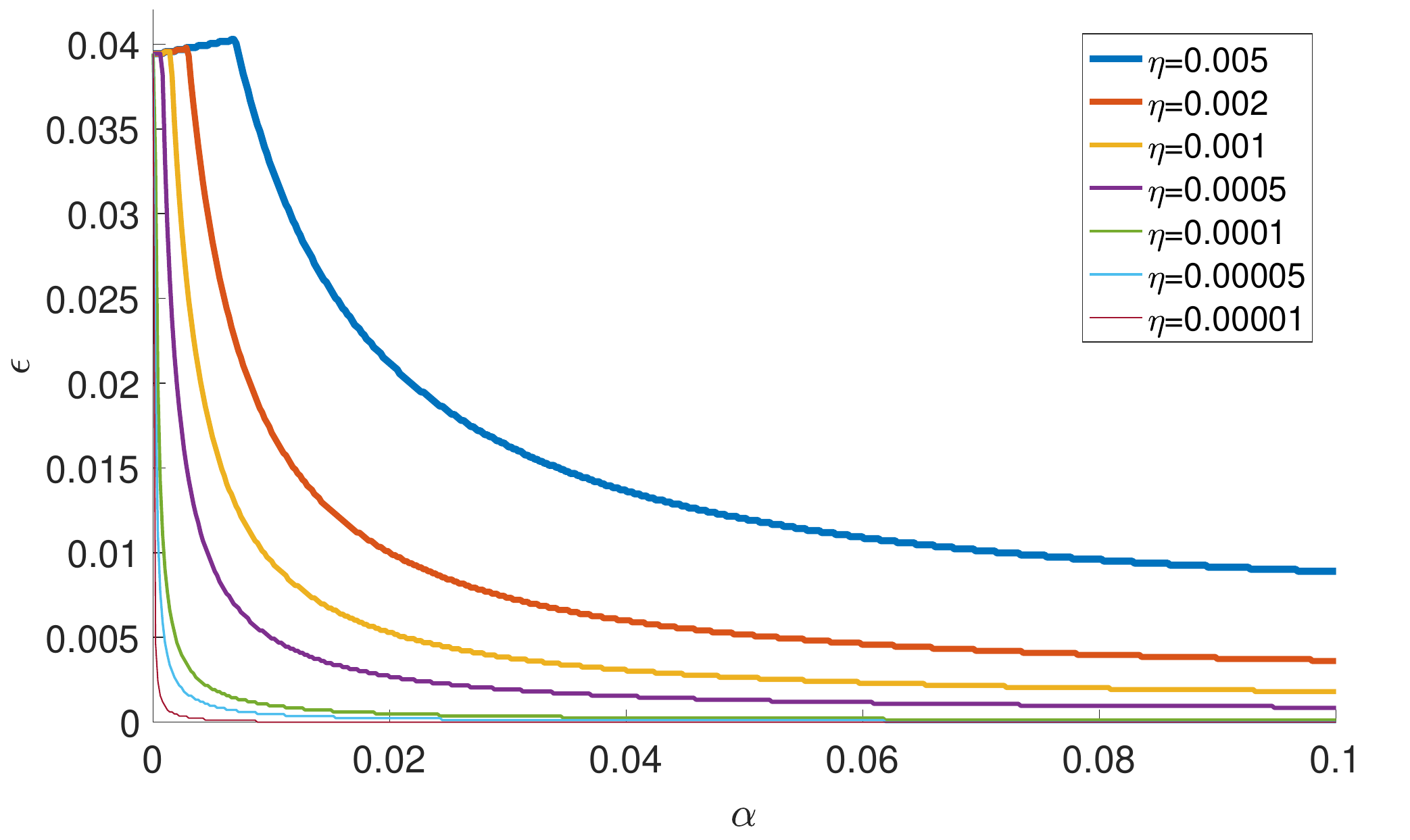}
\caption{$\eta$-thresholds for decoding a $C^\infty(3,6)$ regular LDPC code with $\alpha$-missing wire Gallager B decoding algorithm over BSC($\ep$).}
\label{fig:eta_b}
\end{figure}

One interesting phenomenon shown in Fig.~\ref{fig:eta_b} is that decoding thresholds first increase with increasing decoder missing connection probability. This error enhancement phenomenon is introduced by the missing connections, essentially resulting in a change of choice for threshold $b$ in each iteration to achieve a lower error rate. This SF phenomenon demonstrates that optimization of degree distribution and threshold $b$ in each iteration can be utilized to combat missing connections. A similar SF result shows that the errors introduced in estimating Markov random field models can be partially canceled and benefit end-to-end inference performance \cite{Wainwright2006}.
SF effects due to noise in computational elements, rather than graphical model structure errors as here, have been observed in \cite{ChenVV2014,KarbasiSSV2014,HamiltonAST2014}
and later specifically in LDPC decoders \cite{BrkicIV2016,VasicIBR2015}.

\section{Moving towards Practice}
\label{sec:practice}
Though performance analysis of LDPC decoders with missing connections using density 
evolution is an important topic in coding theory, our eventual goal is to use analytical understanding
for practical system design. 

Towards this end, we first briefly discuss how one can use DE analysis to study sensitivity of codes and decoders, 
so as to give insight into resource allocation over the entire telecommunications system.  In particular
we ask whether more resources should be spent in manufacturing or in operation.
Second, as DE analysis
is an asymptotic approximation of practical finite-length codes, we also perform 
simulations to understand how well the asymptotics describe finite-length code performance. Finally we note that 
increasing the accuracy of semiconductor fabrication by just a small amount requires a significant 
increase in manufacturing cost (which already takes tens of billions of dollars to build facilities, and limits growth of the industry).
As such, we perform preliminary manufacturing yield analysis to show potential industrial impact.
For brevity, this section is largely restricted to Gallager A.

\subsection{Decoder Sensitivity}
\label{sec:sensitivity}
Should the industry invest more resources in operating good communication channels or in manufacturing better receiver hardware?

Taking derivatives of the density evolution function $x_{\ell+1}=f_{DE}(x_l, \ep, \al)$ with respect to $\ep$ and $\al$ and evaluating at $\bar{P_e}^{(\infty)}=x_{\ell}=x_{\ell+1}$, we find the impact of channel noise level and missing connection probability on the final error rate.
\begin{align*}
\bar{P_e}^{(\infty)}(\ep,\al)&=\ep\lambda(\al)+\lambda \left( (1-\ep)p^+(\bar{P_e}^{(\infty)}(\ep,\al)) \right. \\
&\left.\quad +\ep p^-(\bar{P_e}^{(\infty)}(\ep,\al))  \right).
\end{align*}
Denote $$g(x)=(1-\ep)p^+(x(\ep,\al))+\ep p^-(x(\ep,\al)).$$
Take partial derivatives of each side with respect to $\ep$:
\begin{align*}
\tfrac{\partial x(\ep,\al)}{\partial \ep}&=\lambda(\al)
+\tfrac{\partial \lambda \left(g(x(\ep,\al))\right)}{\partial x(\ep,\al)}\tfrac{\partial x(\ep,\al)}{\partial \ep} +\tfrac{\partial \lambda(g(x(\ep,\al)))}{\partial \ep}\\
&=\frac{ \lambda(\al)+\frac{\partial \lambda(g(x(\ep,\al)))}{\partial \ep}}{ 1- \frac{\partial \lambda(g(x(\ep,\al)))}{\partial x(\ep,\al)}}.
\end{align*}
Similarly,
\begin{align*}
\tfrac{\partial x(\ep,\al)}{\partial \al}&=\ep\tfrac{\partial \lambda(\al)}{\partial \al}+\tfrac{\partial \lambda(g(x(\ep,\al)))}{\partial x(\ep,\al)}\tfrac{\partial x(\ep,\al)}{\partial \al}+\tfrac{\partial \lambda(g(x(\ep,\al)))}{\partial \al}\\
&= \frac{ \ep\frac{\partial \lambda(\al)}{\partial \al}+\frac{\partial \lambda(g(x(\ep,\al)))}{\partial \al}}{ 1- \frac{\partial \lambda(g(x(\ep,\al)))}{\partial x(\ep,\al)}}.
\end{align*}

\begin{figure}
\centering
\includegraphics[width=3.3in, height=2.2in]{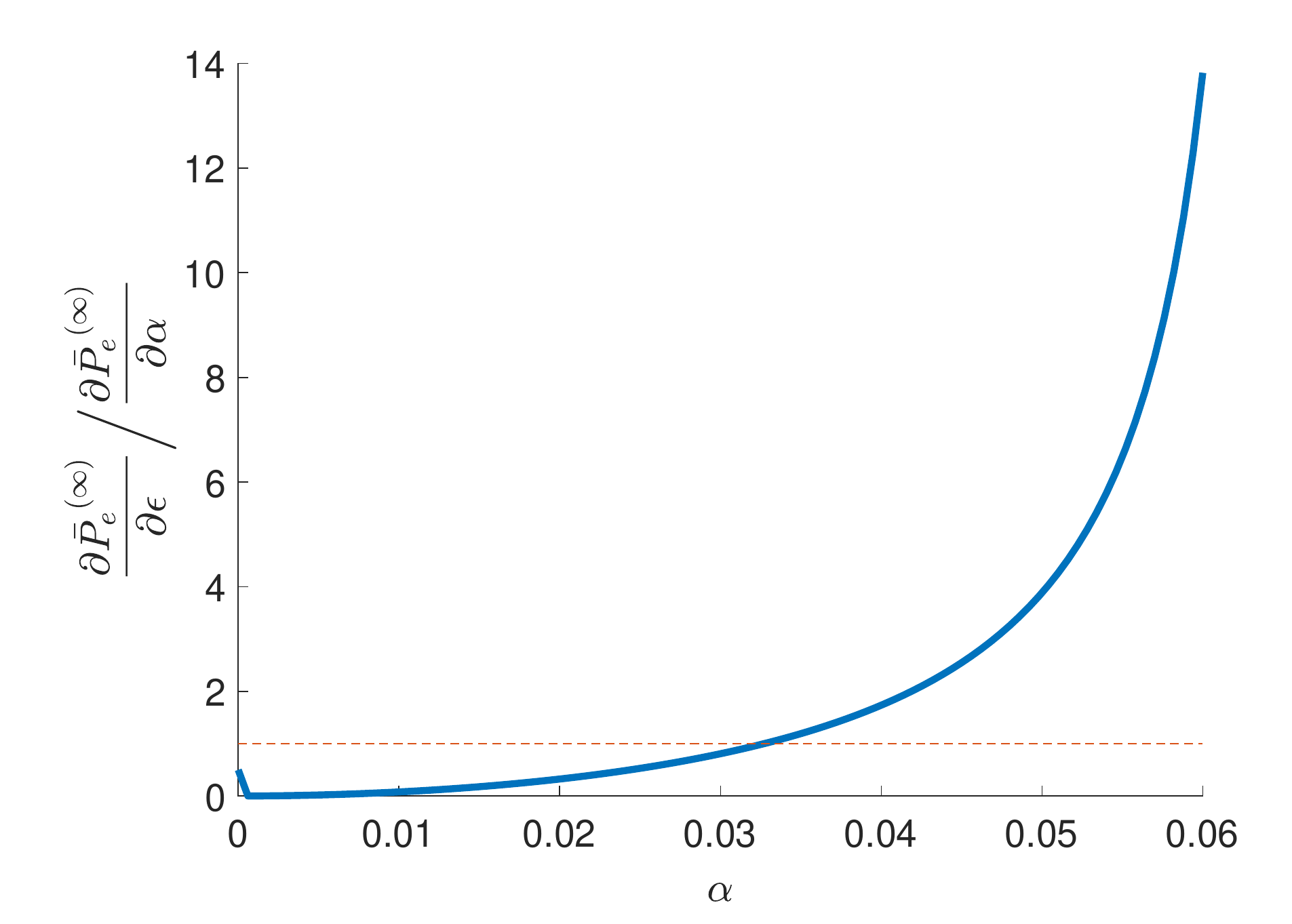}
\caption{Comparison between the derivative of $\bar{P_e}^{(\infty)}(\ep,\al)$ with respect to $\ep$ and $\al$ of decoding a $C^\infty(3,6)$ regular LDPC code with $\alpha$-missing wire Gallager A decoding algorithm over BSC($\ep$), when $\ep$ and $\al$ are at the boundary of decoder useful region.}
\label{fig:sens}
\end{figure}

Fig.~\ref{fig:sens} illustrates the ratio of the derivative of $\bar{P_e}^{(\infty)}(\ep,\al)$ with respect to $\ep$ and $\al$, when $\al$ and $\ep$ are at the boundary of useful region depicted in Fig.~\ref{fig:thresbsc}:
$$\frac{\partial \bar{P_e}^{(\infty)}(\ep,\al)}{\partial \ep}\bigg/ \frac{\partial \bar{P_e}^{(\infty)}(\ep,\al)}{\partial \al}.$$
Different from our intuition, both derivate values are negative at the boundary of the useful region. Recall the linear relationship of $\ep$ and $\al$ at the boundary of the useful region; with the increase of $\al$, $\ep$ has to decrease in order to stay in the useful region, resulting in the decrease in $\bar{P_e}^{(\infty)}(\ep,\al)$.

When operating at the edge of the useful region, as we can see in Fig.~\ref{fig:sens}, it is advantageous to put resources into circuit manufacturing up to an $\alpha$ value of roughly $0.03$ where the curve crosses the equal-ratio point, whereas it is advantageous to put resources into the channel thereafter.  Thus aiming for manufacturing that achieves such a crossover point $\alpha$ may be an appropriate resource allocation strategy.

\subsection{Finite-Length Simulation}
We simulate finite-length systems having decoders with either transiently or permanently missing connections, to demonstrate performance is comparable in the two settings and predicted by density evolution. For $(3,6)$-regular LDPC codes with blocklength $n=498$, $1002$, and $1998$ drawn at random from the code ensemble using socket-switching, we randomly simulate decoding performance with connections either permanently removed before the decoding starts or transiently removed during each decoding iteration for various sets of $(\ep, \al)$. For each trial, decoding is performed for more than 30 iterations (error probability usually convergences within 10 iterations in fault-free decoding). For each channel noise level and missing connection probability, the decoding error probability is averaged over $100$ randomly selected code realizations and missing connection realizations.

\begin{figure}
\centering
\includegraphics[width=3.5in, height=2.2in]{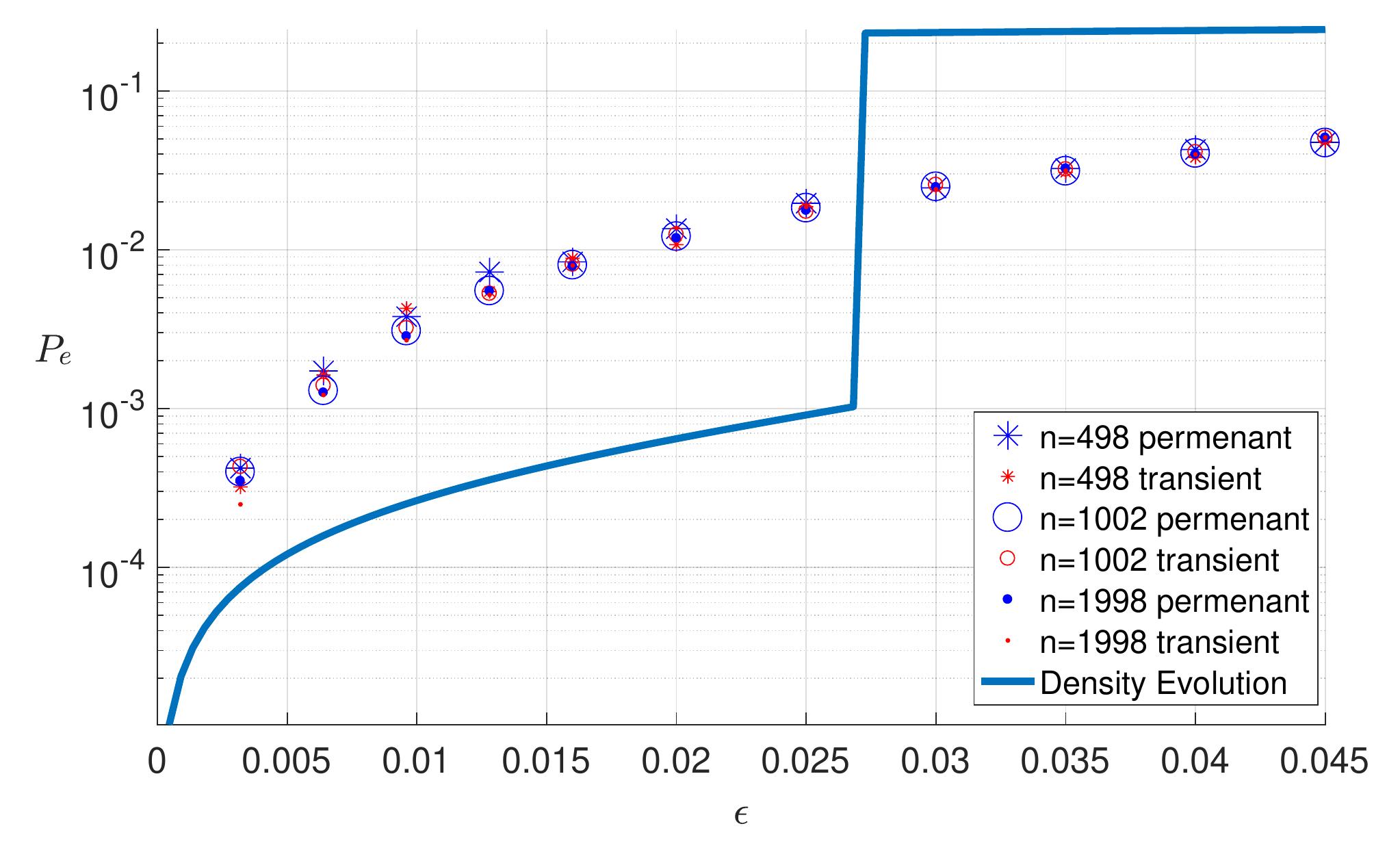}
\caption{$P_e$ of decoding $(3,6)$ LDPC code with finite block-length under Gallager A decoding algorithm over BSC with permanently and transiently missing connection probability $\al=0.02$.}
\label{fig:finite}
\end{figure}

As Fig.~\ref{fig:finite} illustrates, the performance of finite-length codes resembles the asymptotic performance of codes. As expected, below the decoding threshold, the symbol error rates of finite-length codes are higher than asymptotic performance. For fault-free decoders with channel noise below threshold, the asymptotic symbol error rate is $0$, whereas in the case of finite-length codes $P_e$ increases smoothly with increasing $\ep$ \cite{AmraouiMRU2009}. Similar to fault-free decoders for finite-length codes, decoders with missing connections show a similar trend of increasing $P_e$. Further, notice that the performances of transiently and permanently missing connection cases are close to one another. We chose $\al=0.02$ because it is within the defective interconnect range of $1$--$15$\% \cite{HaselmanH2010}; see also Sec.~\ref{sec:sensitivity}. Different from \cite{AmraouiMRU2009}, we do not expurgate codes with small stopping sets. Also recall from Thm.~\ref{thm:conc} that the concentration of the individual performance around the ensemble average is exponential in blocklength and the concentration happens more slowly in the case of missing connections compared to fault-free decoders. Hence, there is more numerical variation in the simulation results at all blocklengths, especially for small $n$. Nevertheless, simulations show that the asymptotic analysis of decoders with missing connections has practical significance.

\subsection{Semiconductor Manufacturing Yield Analysis}
By understanding the computational purpose of circuits (here decoding) it is often possible to raise effective manufacturing yield above the raw yield where all components must be fault-free \cite{BreuerZ2008,JiangG2009}.
To demonstrate that the effective yield of LDPC decoder circuits increases by allowing missing connection probability $\al$
that still guarantees decoding performance $\eta$, we apply the error-tolerant methodology
\cite{BreuerZ2008,JiangG2009}. Threshold testing in \cite{BreuerZ2008} accepts a chip when the chip's performance passes a specified threshold; the increase in effective yield is the amount of chips with defects but still meeting the performance requirement. For decoders with missing connections, this threshold is the maximum symbol error rate $\eta$. We want to find the highest missing connection probability $\al_{max}(G^n,\ep,\eta)$ such that for every decoder with $\al \le \al_{max}$, the resulting $P_e$ is under the target error rate $\eta$ for a given code ensemble $G^n$ and channel quality $\ep$.

Let $\phi(\al)$ be the yield factor, the expected percentage of decoders with missing
connection probability $\al$, and $p(\al)$ be the probability that the circuit has defect density $\al$, often taken as an
exponential distribution \cite{JiangG2009}. Then the effective yield is: 
$$Y=\int_{0}^{\al_{max}(G^n,\ep,\eta)} p(\al)\phi(\al) d\al.$$
For the $\mathcal{C}^{\infty}(3,6)$ LDPC code ensemble and $\eta=10^{-5}$, for a large range of possible channel values $\ep$, Fig.~\ref{fig:thresbec} shows us that $\alpha_{max}=0.01$ is more than sufficient for the case of peeling decoder under BEC. It is straightforward to see that, compared to the yield of the fault-free case $Y_0= p(0)\phi(0)$, allowing some error-tolerance in manufacturing may increase effective yield significantly.  For the exponential distribution function, the absolute increase in yield is linear in $\al_{max}$ \cite{JiangG2009}:
\[
\Delta Y = Y - Y_0 = \al_{max}\tfrac{D_0 A}{(1 + AD_0)^2}\mbox{,}
\]
where $D_0$ is the defect density (average number of defects per unit of chip area), and $A$ is the chip area.
Likewise the fractional increase in yield, is:
\[
\Delta Y / Y_0 = \tfrac{Y - Y_0}{Y_0} = \al_{max}\tfrac{D_0 A}{1 + AD_0} \mbox{.}
\]
As shown in the previous sections, a small defect rate $\al$ does not degrade the performance too much. However, as reported in the semiconductor manufacturing industry, a 1\% reduction in yield can result in a 12\% reduction in profit \cite{Flamm2010, Mittal2016}. Hence even allowing a small probability of defects $\al$ can save a significant amount of wastage and cost without much change in performance.

\section{Conclusion}
\label{sec:clus}
This paper investigated the performance of message-passing decoders  with  transiently  and  permanently  missing  connections  that  might  be  caused  by  process  variation  in  manufacturing  or  timing  errors  in  intra-chip  communications (or both). We derived density evolution equations to characterize the error probability in the peeling decoder over the BEC and modifications of the Gallager A and Gallager B decoders over the BSC, using erasure symbols to represent missing connections. Although the error probability cannot be driven to $0$ in the presence of missing connections, it can be suppressed to a small value $\eta$ when the channel noise level is under a certain decoding threshold $\ep^*$. That is, $\eta$-reliable communication is possible with faulty decoders with missing connections. In a sense, even when the encoder and decoder speak different languages, the result is not catastrophic. A novel structural stochastic facilitation is also observed in Gallager B decoders with missing connections. 

Future work involves considering not just decoders with missing connections, but also miswired and noisy decoders. One may also design new decoder architectures to ensure reliable communication even with miswiring; for example, horizontal connections, a crucial structure in the cortex contributing to the filling in of missing parts in visual images \cite[Ch. 8.33]{Brodal2016}, can be added to decoder designs. Code optimization and new decoding algorithms can also be utilized to take advantage of the stochastic facilitation effect.

\section*{Acknowledgment}
The authors thank A.~Patil and N.~R.\ Shanbhag for discussions on nanoscale circuits, anonymous reviewers for 
helpful comments, and A.~C.~Singer and J.~Cowan for encouragement. 

\begin{appendices}

\section{Probability Theory Definitions}
\label{app:prob}
Before diving into the proof of Thm.~\ref{thm:conc}, some probability theory definitions and the Hoeffding-Azuma inequality are reviewed here.
Consider a space $(\Omega,\mathcal{F})$, where $\Omega$ is a sample space, and a $\sigma$-algebra $\mathcal{F}$ contains subsets of $\Omega$. A random variable $Z$ is an $\mathcal{F}$-measurable function from a probability space into the real number.  If there is a
collection $(Z_{\gamma} | \gamma \in C)$ of random variables $Z_{\gamma}: \Omega \to \mathbb{R}$, then
\[
\mathcal{Z} = \sigma(Z_{\gamma} | \gamma \in C)
\]
is defined to be the smallest $\sigma$-algebra $\mathcal{Z}$ on $\Omega$ such that
each map $(Z_{\gamma}|\gamma \in C)$ is $\mathcal{Z}$-measurable. 

\begin{defin}[Filtration]
Let $\{\mathcal{F}_i\}$ be a sequence of $\sigma$-algebras with respect to the same 
sample space $\Omega$.  These $\mathcal{F}_i$ are said to form a \emph{filtration}
if $\mathcal{F}_0 \subseteq \mathcal{F}_1 \subseteq \cdots$ are ordered
by refinement in the sense that each subset of $\Omega$ in $\mathcal{F}_i$ is
also in $\mathcal{F}_j$ for $i \le j$.  Also $\mathcal{F}_0 = \{\emptyset,\Omega  \}$.
\end{defin}

The conditional expectation of a random variable $Z$ given a $\sigma$-algebra $\mathcal{F}$ is a random variable denoted by $\E{Z|\mathcal{F}}$.

\begin{defin}[Martingale]
Let $\mathcal{F}_0 \subseteq \mathcal{F}_1 \subseteq \cdots$ be a filtration 
on $\Omega$ and let $Z_0,Z_1,\ldots$ be a sequence of random variables on 
$\Omega$ such that $Z_i$ is $\mathcal{F}_i$-measurable.  Then $Z_0,Z_1,\ldots$
is a \emph{Martingale} with respect to the filtration 
$\mathcal{F}_0 \subseteq \mathcal{F}_1 \subseteq \cdots$ if 
$\E{Z_i|\mathcal{F}_{i-1}} = Z_{i-1}$.
\end{defin}
\begin{defin}[Doob's Martingale]
Let $\mathcal{F}_0 \subseteq \mathcal{F}_1 \subseteq \cdots$ be a filtration 
on $\Omega$ and let $Z$ be a random variable on $\Omega$.  Then the sequence
of random variables $Z_0,Z_1,\ldots$ such that $Z_i = \E{Z|\mathcal{F}_i}$
is a Doob's Martingale.
\end{defin}

\begin{lemma}[Hoeffding-Azuma Inequality \cite{Azuma1967,Hoeffding1963,RichardsonU2001}]
\label{lem:azuma}
Let $Z_0,Z_1,\ldots$ be a Martingale with respect to the filtration 
$\mathcal{F}_0 \subseteq \mathcal{F}_1 \subseteq \cdots$ such that
for each $i > 0$, the following bounded difference condition is satisfied
\[
|Z_i - Z_{i-1}| \le \alpha_i \mbox{, }\alpha_i \in [0,\infty)\mbox{.}
\]
Then for all $n > 0$ and any $\xi > 0$,
\[
\Pr\left[|Z_n - Z_0| \ge \xi\right] \le 2\exp\left( -\frac{\xi^2}{2\sum_{k=1}^n \alpha_k^2} \right) \mbox{.}
\]
\end{lemma}

\section{Concentration: Permanently Missing Connections}
\label{app:concp}
The proof of Thm.~\ref{thm:conc} is an extension from and largely identical to \cite[Thm. 2]{Varshney2011}, \cite[Thm. 2]{RichardsonU2001}, or \cite[Thm. 4.94]{RichardsonU2008}. We want to construct a Doob's Martingale with respect to the fraction of error held on each edge during the random revealing process and to show that the difference of the object of interest between each iteration is bounded by a number not related to the number of iterations.

Recall $Z$ denotes the number of incorrect values held at the end of the $\ell$th iteration for a specific $(g, y, w) \in \Omega$, where $g$ is a specific bipartite Tanner graph to represent the choice of LDPC code with variable node degree $\dv$ and check node degree $\dc$, $y$ is a specific input to the decoder, $w$ is a particular realization of the decoder with missing wires, and $\Omega$ is the sample space. Let $\equiv_i$, $0\le i\le m$ be a sequence of equivalence relations on $\Omega$ ordered by refinement, such that $(g',y',w') \equiv_i (g'',y'',w'')$ implies $(g',y',w') \equiv_{i-1} (g'',y'',w'')$. The equivalence relations define equivalence classes by partial equalities such that $(g',y',w',u') \equiv_i (g'',y'',w'',u'')$ \emph{if
and only if} the realizations of random quantities revealed in the first $i$ steps for both pairs is the same.  

Next we use the technique of exposing the edges in the decoding graph in sequence. The first case is when wires are permanently missing. Note that even with positive probability of missing connections $\al$, for a specific code realization, the number of potentially connected edges can be at most $n\dv$. Hence, we expose at most $n\dv$ edges one at a time. At step $i \le n\dv$, we expose the particular check node socket that is connected to the $i$th variable node socket. Next, in the following $n$ steps, we expose the received values $y_i$ from the channel one at a time. At the end of the $n(\dv+1)$ steps, the decoder missing wire probability is also realized, since the defect is permanent. Then we have $(g',y',w') \equiv_i (g'',y'',w'')$ if and only if the information revealed in the first $i$ steps for both pairs is the same.

Now, define $Z_0, Z_1,..., Z_m$ by
$$Z_i(g,y,w)= \E{Z(g',y',w')| (g',y',w') \equiv_i (g,y,w)},$$
where $Z_0=\E{Z}$ and $Z_m=Z$. By construction, $Z_0, Z_1,..., Z_m$ is a Doob's Martingale.
We then use Lem.~\ref{lem:azuma} to give bounds on
$$\Pr[|Z-\E{Z}|>n \dv \ep/2]=\Pr[|Z_m-Z_0|>n \dv \ep/2].$$
To use Azuma's inequality, we first need to prove that for each consecutive member in the sequence $Z_0, Z_1,..., Z_m$, the difference is bounded:
$$|Z_{i+1}(g,y,w)-Z_i(g,y,w)|\le \dl_i, i=0,1,...,m-1$$
where $\dl_i$ depends on $\dv, \dc, \text{and } \ell.$

It was shown in \cite{RichardsonU2001} that for the fault-free decoder without any missing wire, when edges are exposed,
$$|Z_{i+1}(g,y,w)-Z_i(g,y,w)|\le 8(\dv \dc)^{\ell}, 0 \le i \le n\dv.$$
In our case when there exist permanently missing connections, the difference when exposing edges is that the number of edges existing is smaller, and bounded by $n\dv$. The expected number of edges left is $n\dv(1-\al)$. The bound established above still holds with a change of the steps number:
$$|Z_{i+1}(g,y,w)-Z_i(g,y,w)|\le 8(\dv \dc)^{\ell}, 0 \le i \le n\dv.$$
It was also shown that when channel outputs are revealed, the difference in each element in the sequence is bounded by
$$|Z_{i+1}(g,y,w)-Z_i(g,y,w)|\le 2(\dv \dc)^{\ell},$$
where $n\dv \le i \le n(\dv+1)$ in the case where some wires are permanently missing.
Then the theorem follows from applying Azuma's inequality to the Martingale constructed.

\section{Concentration: Transiently Missing Connections}
\label{app:conct}
The second case is when wires are transiently missing at each decoding iteration. The Martingale is constructed differently. Instead of exposing edges, at $\ell$ iterations, we sequentially expose the realization of edges at different iterations. Since each edge is missing independently from others with probability $\al$, only sockets whose nodes are connected through these edges are affected. In each iteration, there are two realizations for each edge (present or missing), then for all previous $\ell$ iterations, the total number affected edges is bounded by $2(2\dv \dc)^{\ell}$. With symmetry of switching node sockets:
$$|Z_{i+1}(g,y,w)-Z_i(g,y,w)|\le 8(2\dv \dc)^{\ell}$$ 
where $n(\dv+1) \le i \le m$.

Hence, in the transiently missing wire case, the bounded difference $\dl_i=8(2\dv \dc)^{\ell}$. The theorem follows from applying Azuma's inequality to the Martingale constructed.
\end{appendices}

\bibliographystyle{IEEEtran} 
\bibliography{abrv,conf_abrv,lrv_lib}

\begin{IEEEbiography}[{\includegraphics[width=1in,height=1.25in,clip,keepaspectratio]{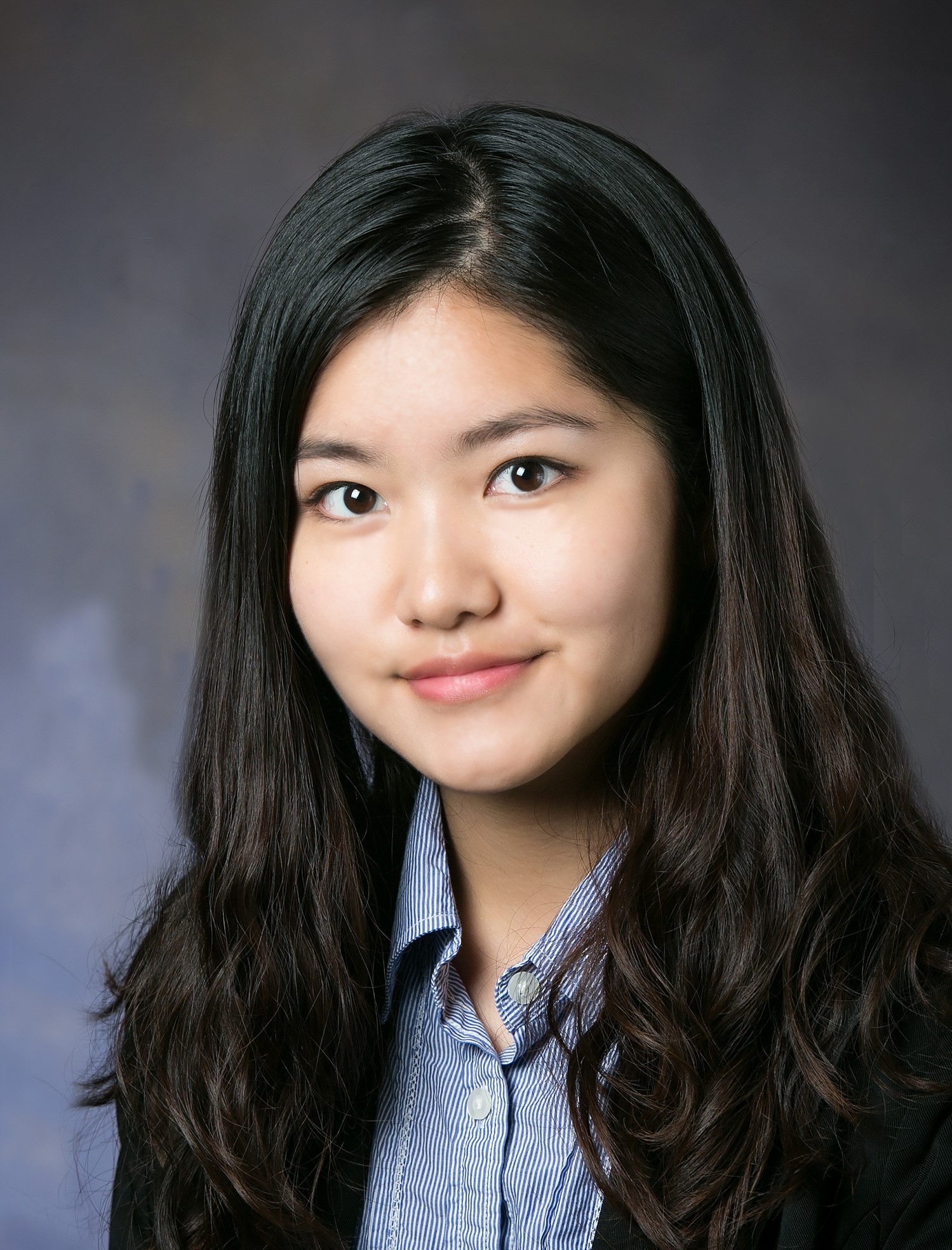}}]{Linjia~Chang}(S'14)
received the B.S.~degree with honors and the M.S.~degree in electrical and computer engineering from the University of Illinois at Urbana-Champaign in 2014 and 2016 respectively. Her research interests span stochastic information processing systems, coding theory, networks, and data analytics. She is a member of Eta Kappa Nu.
\end{IEEEbiography}

\begin{IEEEbiographynophoto}{Avhishek~Chatterjee}
received the Ph.D.~degree in electrical and computer engineering from the University of Texas at Austin in 2015. He is currently a post doctoral research associate with the Coordinate Science Laboratory, University of Illinois at Urbana-Champaign. His research interests lie in theoretical studies of dynamics, optimal designs, and operations of stochastic networks. He is involved in dynamics and inference in social networks, fundamental limits and optimal operations of crowdsourcing systems, resource allocation and dynamics in communication and computer networks, and fundamental limits and resource allocation in nanoscale circuits.
\end{IEEEbiographynophoto}

\begin{IEEEbiography}[{\includegraphics[width=1in,height=1.25in,clip,keepaspectratio]{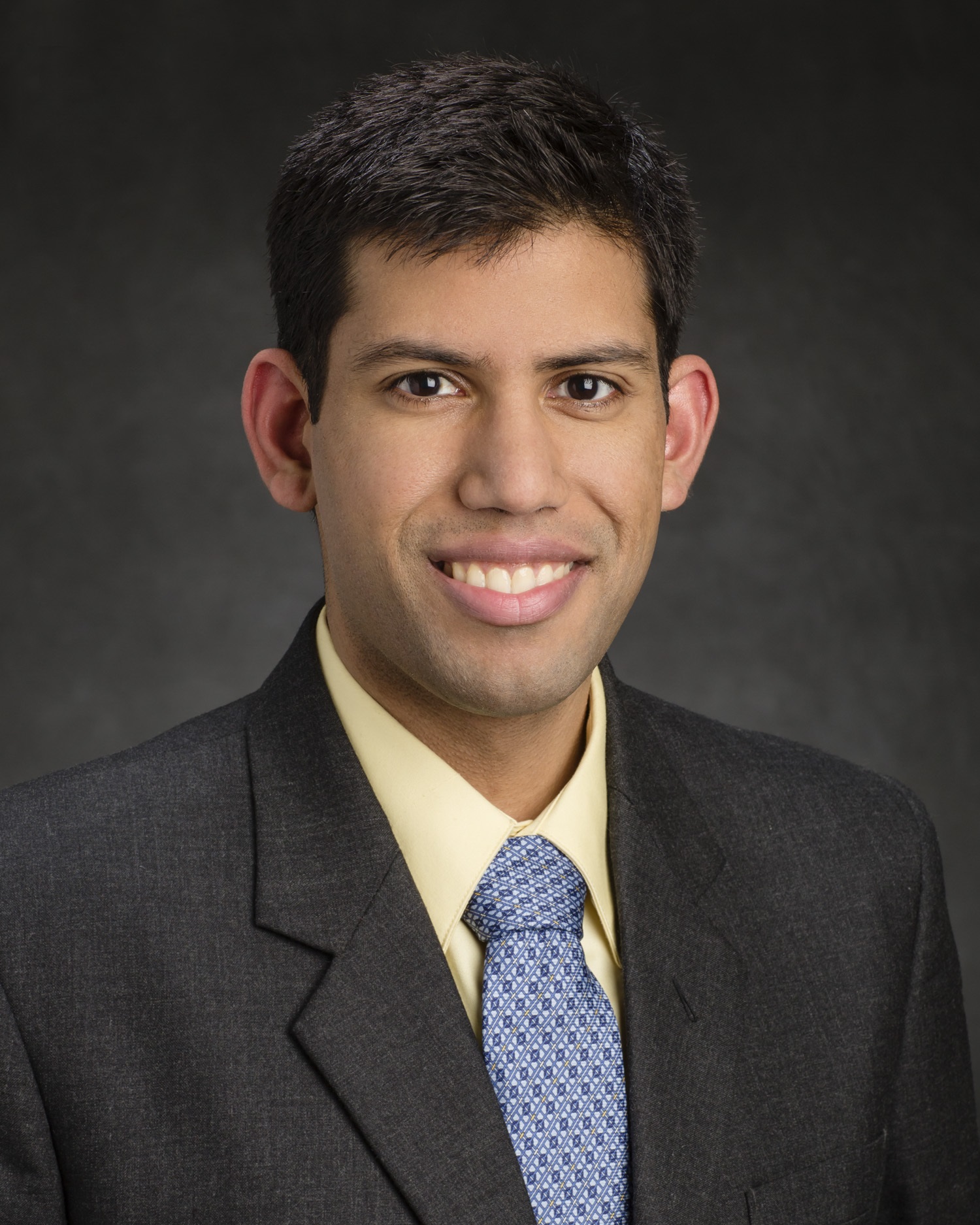}}]{Lav~R.~Varshney}
(S'00--M'10--SM'15) received the B.S.~degree (\emph{magna cum laude}) in electrical and computer engineering with honors from Cornell University, Ithaca, New York, in 2004. He received the S.M., E.E., and Ph.D.\ degrees, all in electrical engineering and computer science, from the Massachusetts Institute of Technology, Cambridge, in 2006, 2008, and 2010, where his theses received the E.~A.\ Guillemin Thesis Award and the J.-A. Kong Award Honorable Mention.

He is an assistant professor in the Department of Electrical and Computer Engineering, the Coordinated Science Laboratory, the Beckman Institute, and the Neuroscience Program at the University
of Illinois at Urbana-Champaign. During 2010--2013, he was a research staff member at the IBM Thomas J.\ Watson Research Center, Yorktown Heights, New York. His research interests include information and coding theory; limits of nanoscale, human, and neural computing; human decision making and collective intelligence; and creativity.

Dr.\ Varshney is a member of Eta Kappa Nu, Tau Beta Pi, and Sigma Xi. He
received the IBM Faculty Award in 2014 and was a Finalist for the Bell Labs Prize,
in 2014 and 2016.  He and his students have won several best paper awards.  His work appears in the anthology, \emph{The Best Writing on Mathematics 2014} (Princeton University Press). He is a founding member of the  IEEE Special Interest Group on Big Data in Signal Processing and currently serves on the Shannon Centenary Committee of the IEEE Information Theory Society.  He also currently serves on the advisory board of the AI XPRIZE.
\end{IEEEbiography}

\end{document}